\documentclass[%
 reprint,
 amsmath,amssymb,
 aps,
]{revtex4}

\usepackage[T1]{fontenc} 

\usepackage{amssymb}   
\usepackage{amsmath}   

\usepackage{graphicx}
\usepackage{subfigure}
\usepackage{hyperref}
\usepackage{float}
\usepackage{color}

\usepackage{graphicx}
\usepackage{dcolumn}
\usepackage{bm}

\begin{document}


\title{\boldmath Assessing the impact of bulk and shear viscosities on large scale structure formation}

\author{C.M.S. Barbosa}
 \altaffiliation{carolina.barbosa@aluno.ufes.br}
\author{H. Velten}%
 \email{velten@pq.cnpq.br}
\affiliation{Universidade Federal do Esp\'irito Santo, \\Av. Fernando Ferrari, Goiabeiras, Vit\'oria, Brasil}
\author{J. C. Fabris}%
 \email{fabris@pq.cnpq.br}
\affiliation{Universidade Federal do Esp\'irito Santo, \\Av. Fernando Ferrari, Goiabeiras, Vit\'oria, Brasil}

\author{ Rudnei O. Ramos}
 \email{rudnei@uerj.br}
\affiliation{Universidade do Estado do Rio de Janeiro, 
Departamento de F\'{\i}sica Te\'orica,\\
20550-013 Rio de Janeiro, RJ, Brazil
}
\affiliation{
 Physics Department, McGill University, Montreal, QC, H3A 2T8, Canada}


\begin{abstract}
It is analyzed the effects of both bulk and shear viscosities on the perturbations,
relevant for structure formation in late time cosmology.
It is shown that shear viscosity can be as effective as the bulk viscosity 
on suppressing the growth of perturbations and delaying the nonlinear regime.
A statistical analysis of the shear and bulk viscous effects is performed
and some constraints on these viscous effects are given.
\end{abstract}

\pacs{Valid PACS appear here}
\maketitle


\section{Introduction}

The current cosmological concordance model is the so called
$\Lambda$CDM model, which assumes a homogeneous and isotropic Universe,
the gravitational interaction is dictated by general relativity, 
making use of pressureless cold dark matter (CDM) and a cosmological
constant $\Lambda$. It is a successful model since it
fits with high statistical confidence many observational tests, like
the Cosmic Microwave Background (CMB), Supernovae, Baryonic Acoustic
Oscillations (BAO) and indirect estimations for the Hubble parameter as
a function of redshift, $H(z)$. It is called the concordance
model because all such observations can be described using the same
set of parameter values. The $\Lambda$CDM model gives a hydrodynamic treatment
for the matter components of the Universe and the description of its
background dynamics is given by the {}Friedmann equation

\begin{equation}
H^2 (a) = H_0^2 \left( \frac{\Omega_{m0}}{a^3} +
\frac{\Omega_{r0}}{a^4} + \frac{\Omega_k}{a^2} + \Omega_\Lambda
\right), 
\end{equation}
where $a$ is the scale factor, $H_0$ is the present value for the Hubble parameter, 
$\Omega_{m0}$ generically denotes the fractional
matter density components (assumed pressureless) of the Universe,
$\Omega_{r0}$ is the fractional radiation density term, $\Omega_k$ is the term related to
the curvature and $\Omega_\Lambda$ denotes the cosmological constant
component. The last one is essential to describe the late time accelerated
expansion of the Universe. In practice, the contribution of the
radiation at late times (i.e., at the time of structure formation) is
negligible compared to the matter and cosmological constant
terms. Also, observations indicate that the geometry of the Universe is almost flat. Hence the contribution of
the curvature is practically irrelevant, $\Omega_k \approx 0$.

Despite the $\Lambda$CDM model being in agreement with several
observational tests, there are still a few (but important) problems
unsolved. Among the problems with $\Lambda$CDM model are worth citing 
the excessive agglomeration of matter due to the nature of the CDM and 
the puzzle of missing satellites~\cite{Moore:1999nt,Klypin:1999uc}, 
the cusp-core problem~\cite{cuspcore,cuspcore1,cuspcore2} and also the issue related to
the fact that the Planck collaboration observed less clusters than
expected~\cite{Planck}. These problems, in principle unexplained by
the $\Lambda$CDM model, motivate finding possible modifications for the current
cosmological concordance model in ways that could solve such issues.

Given that the standard CDM model is plagued by the apparent excess of
clustered structures, we can see this as a possible clue on the 
role that some physical mechanism able to suppress the density contrast growth can
play in solving the small scale problems.  To address the problem of
finding such suppression mechanism, the approach we will use in this
work is to relax the assumption that dark matter behaves on large
scales as a perfect and adiabatic fluid and, instead, we will assume that
it behaves as a
fluid with natural dissipative effects built in. It has been
previously studied models with the introduction of dissipative effects
in the CDM component as a step towards alleviating the problems above
mentioned, e.g., making use of bulk viscous effects~\cite{zimdahl:1996, 
fabris:2006, zimdahl:2007, piattella:2011}. The
inclusion of a bulk viscosity makes the model more realistic since during
the structure formation era some non-perfect processes ought to occur.

There are many possible dissipative processes that can occur during
the cosmic evolution, such as particle
production~\cite{zeldovich,prigo,ademir,pacheco,Ramos:2014dba,Santos:2014kia},
matter diffusion~\cite{Calogero:2013zba} and fluid
viscosity~\cite{pad,zimdahl:1996}. Within the latter, dissipative
effects on the radiation fluid, such as bulk and shear viscosities,
has been shown to be of particular importance during the inflationary
epoch~\cite{Maartens:1996dk, delCampo:2007cy, BasteroGil:2011xd, BasteroGil:2012zr, 
Bastero-Gil:2014jsa, Giovannini:2015uia}
and also as a possible description for dark matter and dark
energy~\cite{murphy,Kremer:2002hz,Wilson:2006gf,Gagnon:2011id,Floerchinger:2014jsa,Blas:2015tla}. 
The option for a
late time accelerated universe was mentioned well before the direct
evidence from Supernovae observations~\cite{pad}. In fact, bulk
viscosity implies a negative pressure contribution that can
accelerate the universe. Although this effect can lead to a realistic 
mechanism for phantom cosmologies~\cite{Velten:2013qna}, its use as the agent 
playing the role of accelerating the late time expansion has been
revealed problematic~\cite{barrow,Velten:2011bg,piattella:2011,Giovannini:2015uia}, and the presence
of dark energy (e.g. in the form of a cosmological constant) seems to
still be necessary. A recent review of the main results on bulk viscous cosmologies is presented in Ref. 
\cite{Brevik:2017msy}.

In previous works the role played by the bulk viscosity in the linear
structure formation process has been studied~\cite{Velten:2015tya,
  Velten:2014xca, Carames:2014cga, Velten:2013pra, Velten:2013rra,
  Velten:2012uv, Velten:2011bg, HipolitoRicaldi:2010mf,
  HipolitoRicaldi:2009je}.  In the present work, however, we aim also to
assess the role played by the shear viscosity and its combined effect
with the bulk viscosity on the structure formation process. Shear
viscosity has mostly been neglected in these type of studies on the
grounds of not contributing for a homogeneous and isotropic universe,
which is certainly true at the background level but this is not the case
at the level of the perturbations. 
References~\cite{Thomas:2016iav,Kopp:2016mhm} should be mentioned here 
since the impact of shear-like effects on CMB have been investigated, 
but using an effective parametrization for the speed of sound which 
differs from the present work. In this work we seek to determine whether or not
shear viscosity alone, and also in combination with the bulk viscosity,
can have a similar impact on the growth suppression as found for the
bulk viscosity alone. As already mentioned, shear viscosity does not
contribute to an isotropic and homogeneous background. However, it
contributes to the evolution of perturbations, which can be non
negligible, as already shown in the case of early universe cosmology,
in particular during
inflation~\cite{BasteroGil:2011xd,Bastero-Gil:2014jsa}. In this work we will 
perform a similar inspection in the context of the dark
matter models. The behavior of the density contrast will be analyzed
and compared with the standard $\Lambda$CDM model. Moreover, the
results will be tested using the redshift space distortion data.
Other dissipative effects, like for instance heat conduction, are not so relevant in our analysis. Such effects are closely connected to the coupling of the baryonic matter with the photon radiation, which are important either at times closer to recombination, or in the highly nonlinear regime of structure formation at small scales (e.g., galaxy formation) and when it becomes important for modeling astrophysical processes.
In fact, it is well known that 
heat conduction can be as important as shear viscosity when considering their effects on the 
photon-baryon plasma, affecting the baryon acoustic oscillations through 
Silk damping~\cite{Silk:1967kq}. Neither of these regimes where heat conduction
would be of relevance will be treated in this work, which is concerned with the linear regime 
only for the cosmological scalar perturbations and it will refers, therefore, 
to an epoch where neither baryons nor photons (radiation) are the most important component. 
This is very reasonable, since in this regime the most important component of matter
is dark matter and given that we still do not have a proper understanding of its nature,
it is reasonable to describe it beyond the ideal, pressureless fluid approximation,
endowing it with viscous properties.
Nevertheless, there might be also an important contribution from the baryons
even in the linear regime and in which case the baryonic fluid is well described
as a pressureless fluid. Thus, we will also investigate how the inclusion of a separated baryonic sector (during the linear regime of structure formation) might impact our bounds on the dark matter viscosity.

This paper is organized as follows. In section~\ref{sec2} we set the
background model, introducing a dissipative dark matter component.  In
section~\ref{sec3}, a perturbative analysis is performed and the relevant 
expressions required for our analysis are derived. In section~\ref{results}
we present our numerical
results concerning how the density contrast is changed as a function of
the bulk and shear viscosities. We also give a statistical analysis and
we determine the preferable values for the viscosities using the most recent
redshift-space-distortion based $f(z) \sigma_8(z)$ data. 
In section~\ref{baryons} we analyze how the inclusion of baryon can change
our results. In
section~\ref{conclusions} we discuss the results obtained and present
our conclusions.

\section{The background dynamics of the $\Lambda$ viscous CDM model}
\label{sec2}

The present cosmological model that we investigate shares some similarities with
the standard $\Lambda$CDM. However, in our approach dark matter
behaves as a viscous/dissipative component. The general structure of
this model is given by the field equation

\begin{equation}
R_{\mu \nu} - \frac{1}{2}g_{\mu \nu} R - \Lambda g_{\mu \nu} = 8\pi G
T_{\mu \nu},
\end{equation}
where $T_{\mu \nu}$ stands for the energy momentum tensor of the
viscous matter. This tensor possesses both the perfect fluid structure
as well as the possible dissipative effects in the form of bulk viscosity $\xi$
and shear viscosity $\eta$, respectively, such that~\cite{Landau:1971,Weinberg:1972}

\begin{equation}
T^{\mu \nu} = \rho u^{\mu} u^{\nu} - p \left( g^{\mu \nu} - u^{\mu}
u^{\nu} \right) + \Delta T^{\mu \nu},
\end{equation}
where the component $\Delta T^{\mu \nu}$ is the viscous contribution
to the fluid,

\begin{align}
&\Delta T^{\mu \nu} = \eta \left[ u^{\mu ; \nu} + u^{\nu  ; \mu} -
  u^{\rho} \nabla_{\rho} \left( u^{\mu} u^{\nu} \right) \right] \nonumber \\
&+ \left( \xi - \frac{2}{3} \eta \right) \left(  g^{\mu \nu} - u^{\mu} u^{\nu}
\right) \nabla_\rho u^\rho.
\end{align}

{}For simplicity, we set the kinetic pressure to $p=0$. Then, our dark
matter possesses only the viscous pressure

\begin{equation}
p_v = -\xi u^\mu_{; \,\mu}.
\label{pEckart}
\end{equation}

The above viscous pressure will be assumed as the total pressure of the dark matter component, 
but here it is expedient a cautionary remark. It is widely known that in non-equilibrium thermodynamics the 
viscous pressure represents a small correction to the positive defined equilibrium (kinetic) pressure. 
This condition applies both to the non-causal Eckart theory~\cite{eckart}, as in Eq.~(\ref{pEckart}), and to 
the causal M\"uller-Israel-Stewart formalism~\cite{muller,israel,stewart}.

In addition to the viscous effects we consider in this work, there could also be kinetic effects that
could be of importance (e.g. like radiation pressure, or velocity dispersion effects).
As far these effects are concerned in the context of dark matter, we recall that in general viable cold dark 
matter candidates (WIMPs, for instance) are based on particle masses of order $m_{dm} \sim \mathcal{O} ({\rm GeV})$, 
resulting in an almost negligible kinetic pressure. 
However, one can promote a fair estimation on $m_{dm}$ such that the kinetic pressure would be relevant. 
Let us assume that the thermal contribution has the upper limit given by the
CMB temperature (it is equal to this temperature in case of thermal equilibrium of all the cosmic matter components).
Note that here we only want to get an estimate on the
upper bound contribution, so we will assume the unlikely case that
WIMPS would be in equilibrium with the CMB (which is clearly not possible,
otherwise these type of DM would already been detected). This temperature today is $T_0 = 2.35\times10^{-13}$ GeV and the (physical) temperature will scale with the redshift as
$T(z) = (1 + z)T_0$. Thus, the temperature decreases as the universe evolves, while the rest mass contribution remains constant. 
As an upper bound estimation, let us suppose that both 
contributions are equal at a given redshift $z$. Hence this assumption implies,
$m = 3(1 + z)T_0/2$.
With this equality occurring, for instance, at the decoupling time $z_{\rm eq} \simeq 1000$, then
we readily obtain an estimate for the mass as being $m \sim 0.3$ eV.
This is of the order of the estimated neutrino mass. If the equality happens more recently, the value of $m$ is still 
smaller. Indeed, WIMPs are expected to decoupled from the primordial bath much earlier around $ T \sim$ GeV. Hence, in order for the kinetic contribution to be relevant, the mass of dark matter particles must be extremely 
small. In fact, for dark matter candidates that can have very small masses, like axions, or other typical dark matter candidates, 
like WIMPs, are expected to freeze out much earlier in the history of the universe and are decoupled from the standard model particles
well before the decoupling time (we recall that present day cosmological observations tends to strongly constrain any form of 
warm dark matter). {}Furthermore, in the linear regime for the perturbations that we will be focusing in this work,
kinetic effects have been estimate to contribute mostly at the percentage-level 
only~\cite{Piattella:2015nda,Adamek:2017grt}.

Using a FLRW metric the {}Friedmann equation reads

\begin{equation}
H^2 \equiv \left(\frac{\dot a}{a}\right)^2= \frac{8 \pi G}{3} \rho_v + \frac{\Lambda}{3},
\label{friedmann}
\end{equation}
where $G$ is the gravitational Newton constant and $\Lambda$ the cosmological
constant. In the above equation we are assuming a flat Euclidean geometry
and we have neglected the radiation contribution, which is much smaller than the
matter component by the time of interest here, when structures start to form
in the Universe and which occurs much later than radiation domination. 
In addition, we will consider in the following that all the matter
components, represented by the $\rho_v$ term, are endowed with viscous 
properties.
Indeed, if only the linear regime structure formation is considered, which is the case considered in this work, 
it is unnecessary a proper separation between baryons and dark matter. Baryonic fluctuations $\delta \rho_b$ follow the 
dark matter ones $\delta \rho_b/\rho_b \rightarrow \delta_{DM}/\rho_{DM}$ in the linear regime. 
One should also recall that baryons totals about $1/6$ of the present total matter distribution, thus, even if our 
analysis fails in providing the correct estimation on the impact of the viscosities due to neglecting the baryons contribution, 
this is not expected to lead to appreciable changes in our bounds on the viscosities. 
We could also in principle assign for the dark and the baryonic
matters different viscous properties, but as far the objective of the present work is concerned, i.e., 
to assess the relative importance of the bulk and shear viscosities
in the linear structure formation process, this is an unnecessary complication. In any case, the baryonic matter 
is a subdominant component with respect to dark matter. Moreover, in the later phases of the evolution 
of the Universe it exhibits generally the pattern determined by the dark matter component.
An explicit analysis of the effects of the baryonic component on our results will be
performed in section~\ref{baryons}.

By defining the fractional densities $\Omega_v = 8 \pi G \rho_v / (3 H^2_0)$ and
$\Omega_\Lambda= \Lambda / (3 H^2_0)$, where $H_0$ is the present value for the Hubble
parameter, then the {}Friedmann equation~(\ref{friedmann}) becomes

\begin{equation}
H^2 = H_0^2 \left(\Omega_v + \Omega_{\Lambda}\right).
\label{H2}
\end{equation}

Using now the fluid equation for $\rho_v$,
\begin{equation}
\dot{\rho}_v + 3 H \left(\rho_v + p_v \right) =0,
\label{dotrho}
\end{equation}
and recalling that for a bulk viscous matter fluid in a FLRW metric the pressure
is $p_v = - 3 H \xi$, we can recast Eq.~(\ref{dotrho}) as an equation for 
the fractional density $\Omega_v$ as
\begin{equation}
a\frac{d \Omega_v}{da} + 3\Omega_v (1 + \omega_v) =0,
\label{dOmegav}
\end{equation}
where we have defined the fluid equation of state parameter for the viscous dark matter fluid,
$\omega_v$, as
\begin{equation}
\omega_v \equiv \frac{p_v}{\rho_v} = -\frac{3H\xi}{\rho_v}.
\label{omegav}
\end{equation}

We note that in the present work the formalism developed in Refs.~\cite{eckart,Landau:1971} 
is employed such as to keep contact with similar previous works. Moreover, for the late Universe (the period of 
the cosmic evolution we are interested in) this non-causal formalism constitutes a good approximation. 
{}Furthermore, the validity of the hydrodynamic equations describing the bulk, and similarly for the shear, 
viscous contributions considered here typically requires $|\omega_v| \ll 1$, otherwise a higher
order hydrodynamics formulation is required~\cite{Hiscock:1983zz} (see for instance Ref.~\cite{BasteroGil:2012zr}
for a contrast of different viscous
hydrodynamic formulations as far as early Universe cosmology is concerned). We will then limit our analysis
within the regime of validity of the present hydrodynamics formalism.

In the following, it will also be useful to define dimensionless bulk and shear viscosities,
$\tilde{\xi}= 24 \pi G \xi/H_0$ and $\tilde{\eta}=24 \pi G \eta/H_0$, respectively.
We will also assume a general form for the viscous coefficients such that~\cite{graves},

\begin{eqnarray}
&&\xi \equiv  (\Omega_v/\Omega_{v0})^{\nu}\, \xi_{0},
\label{xi}
\\  
&&\eta \equiv 
(\Omega_{v}/\Omega_{v0})^{\lambda}\, \eta_0,
\label{eta}
\end{eqnarray}
where the exponents $\nu$ and $\lambda$ are real numbers, 
while $\xi_0$ and $\eta_0$ are constant parameters. 
In general the coefficients of bulk and shear viscosities can be proportional also to the particle free mean path, but this requires the knowledge of the microscopic details of the interactions involving the dark matter particles. 
In this work we are not considering specific candidates for dark matter particles and by assuming the functional forms given by  
Eqs.~(\ref{xi}) and (\ref{eta})  has the advantage of allowing for
a completely model independent analysis. 
In fact, the
above general forms assumed for the viscosities are
quite natural for an isotropic and homogeneous Universe and, under this symmetric 
configuration, it may cover for example the expressions
displayed in Refs.~\cite{Weinberg:1971,Weinberg:1972}.


\section{Perturbative dynamics}
\label{sec3}

To study the perturbative dynamics when including the viscosities, we will work in the Newtonian 
gauge. Hence, the line element for scalar perturbations in an
homogeneous and isotropic flat Universe is

\begin{equation}\label{metric}
ds^2 = a^2(\tau) \left[ -(1+2\phi)d\tau^2 + (1-2\psi)\delta_{ij}dx^i
  dx^j \right],
\end{equation}
where $\tau$ is the conformal time and $\phi$ and $\psi$ are the metric perturbations, which 
are in general equal in the absence of anisotropic stresses, e.g., shear viscosity, but for
dissipative processes, as it will be considered here, they are independent
functions~\cite{brand}.

\subsection{Perturbed Einstein equations}
	
Applying Eq.~(\ref{metric}) to the Einstein equations we obtain, for
example, the $(0,0)$-component, in momentum space, as given by

\begin{equation}\label{Einstein0}
-k^2 \psi - 3\mathcal{H} \left( \psi^{\prime} +  \mathcal{H} \phi
\right) = \frac{3}{2} \Omega_v \mathcal{H}_0^2 a^2 \Delta,
\end{equation}
while the $(0,i)$-component is
\begin{equation}\label{0i}
-k^2 \left( \psi^{\prime} + \mathcal{H}\phi \right) =
\frac{3}{2}\mathcal{H}_0^2 \Omega_v \left( 1 + w_v \right) a \theta,
\end{equation}
where $\mathcal{H}=\frac{a^{\prime}}{a}$, with the symbol "$\prime$ "
corresponding to a derivative with respect to the conformal time,
$k$ is the (comoving) momentum. 
In the above equations we have also defined the density contrast, $\Delta = \delta \rho / \rho$. 
{}From the ($0,i$)-component of the Einstein's equation~(\ref{0i}),
we obtain the definition for the velocity potential $\theta= \partial_i \delta u^{i}$.

{}Finally, the evolution of the potentials $\psi$ and $\phi$ are encoded in the  
$i - j$ component of the Einstein equation,

\begin{align}
& \left[ \psi^{\prime \prime} + \mathcal{H} \left( 2\psi + \phi \right)^{\prime} + 
\left( 2 \mathcal{H}^{\prime} + \mathcal{H}^2 \right) \phi + \frac{1}{2} \nabla \left( \phi - \psi \right) \right] \delta^i_j \nonumber \\ 
& - \frac{1}{2} \partial_i \partial_j  \left( \phi - \psi  \right) = 4 \pi G a^2 \delta T^i_j ,
\end{align}
where 

\begin{eqnarray}
\delta T^i_j &=& \delta p \delta^i_j - \xi \left( \delta u^m_{\, ,m} - 
\frac{3 \mathcal{H}}{a}\phi - \frac{3 \psi^{\prime}}{a} \right) \delta^i_j - 
\left(\delta \xi \right) \frac{3 \mathcal{H}}{a}\delta^i_j \nonumber \\
& -& \eta g^{ik} \delta^l_j \left( \delta u_{k,l} + \delta u_{l,k} - \frac{2}{3}a^2 \delta u^m_{\, ,m} \delta_{kl} \right),
\end{eqnarray}
and from the $i \neq j$ case of the above equation we find

\begin{equation}\label{Einstein1}
-\frac{k^2}{2}(\phi - \psi) = \frac{3\mathcal{H}^2}{\rho}\eta \, \theta.
\end{equation}

{}From Eq.~(\ref{Einstein1}) one notices that $\phi \neq \psi$ if
$\eta \neq 0$. This demonstrates a clear feature of the presence of
shear viscosity (anisotropic stress), i.e., the Newtonian potentials
do not coincide. It is worth noting that $\phi \neq \psi$ is also seem
as a manifestation of modified gravity theories \cite{MGpotentials, MGpotentials1, MGpotentials2,
MGpotentials3}. 
Therefore, this aspect represents an important degeneracy in cosmological perturbation
theory, which is not frequently noticed in the literature. 
 
\subsection{Quasi-static approximation}

We are interested in the evolution of small scale (inside the Hubble
radius) perturbations along the matter dominated period. In such
situation, the Newtonian potentials $\phi$ and $\psi$ almost do not
vary in time. Hence, a quasi-static approximation for these quantities is 
a good approximation. Thus, the continuity equation can be written as

\begin{equation}\label{continuityapprox}
\Delta^{\prime} - 3\mathcal{H} \omega_v\Delta + (1+2\omega_v)(a\theta)
- \frac{9 \mathcal{H}^2 (\delta \xi)}{\rho a} \approx 0,
\end{equation}
and the (0-0) component of the Einstein's equation becomes

\begin{equation}\label{einsteinapprox}
-k^2 \psi \approx \frac{3}{2}\Omega_v \mathcal{H}_0^2 a^2 \Delta.
\end{equation}

By combining Eqs.~(\ref{continuityapprox}) and (\ref{einsteinapprox}), the (0 - i) component equation (\ref{0i}) becomes, 

\begin{align}\label{conservacao_nui}
&(a \theta)^{\prime} + \left[ \mathcal{H} \left( 1 - 3w_v \right) +
    \frac{w_v^{\prime}}{1 + w_v} + \frac{k^2 \left(\tilde{\xi} +
      \frac{4}{3}\tilde{\eta}\right) }{a \rho (1 + w_v)} \right]
  (a\theta) \nonumber \\ & + \frac{k^2 w_v
    \psi^{\prime}}{\mathcal{H}(1+ w_v)} - \frac{k^2\phi}{1+w_v} -
  \frac{w_v k^2 (\delta \xi)}{\xi(1 + w_v)}= 0.
\end{align}

Using the Eqs.~(\ref{H2}), (\ref{dOmegav}) and (\ref{omegav}), we obtain that
\begin{equation}
\frac{a}{H} \frac{dH}{da} = - \frac{3}{2} (1+ \omega_v)\Omega_v \frac{H_0^2}{H^2},
\end{equation}
and
\begin{equation}
a \frac{d \omega_v}{d a} = 3 \omega_v (1+\omega_v) \left(1-\nu -\frac{\Omega_v}{2} \frac{H_0^2}{H^2} \right).
\end{equation}
Then, using Eq.~(\ref{conservacao_nui}), after some algebra, we can express
the equation~(\ref{continuityapprox}) for the density contrast as 
a closed second-order differential equation in the form
\begin{align}\label{euler}
&a^2 \frac{d^2 \Delta}{d a^2}+\left( 3 -\frac{3}{2} \Omega_v \frac{H_0^2}{H^2} +A +
    k^2 B \right) a \frac{d \Delta}{da} \nonumber \\ 
    &+ \left( C + k^2 D \right) \Delta = 0,
\end{align}
where the factors $A$, $B$, $C$ and $D$ appearing in the above equation 
are defined, respectively, as

\begin{align}
&A = \frac{3 \omega_v}{1+2 \omega_v} \left[ 2 \nu (1+\omega_v) - 3 - 4 \omega_v-\omega_v \Omega_v 
\frac{H_0^2}{H^2}\right] \label{termA}
\nonumber \\
&- \frac{2\omega_v}{1+\omega_v} \frac{R}{\Omega_v}, \\
& B = -\frac{w_v (1+\frac{4}{3}R)}{3 H^2 a^2 (1+ w_v)}, \label{termB} \\
& C = -\frac{3\Omega_v}{2}\frac{H_0^2}{H^2} \Big{\lbrace}  \frac{1}{(1+\omega_v)(1+2\omega_v)}+ \frac{\omega_v}{1+2\omega_v} [3 \nu(4 \omega_v^2 \nonumber \\
&+5\omega_v+2) -12 \omega_v^2-15\omega_v-2]\Big{\rbrace} - \frac{3 (1-\nu)\omega_v}{1+2\omega_v} [-3\nu (2\omega_v^2  \nonumber \\
&+2 \omega_v+1) + 7 \omega_v+5 ] +\frac{2(1-3\nu) \omega_v^2}{1+\omega_v} R, \label{termD} \\
& D = \frac{ w_v^2 (1+\frac{4}{3}R)}{ H^2 a^2 (1+ w_v)}(1-\nu) +
  \frac{\nu \omega_v \left( 1 + 2w_v \right)}{1+w_v}\left(
  \frac{\Omega_v}{\Omega_{v0}} \right)^\nu
\label{termD}
\end{align}
where in the above equations we have also introduced the quantity $R\equiv\tilde{\eta}/\tilde{\xi}$,
i.e., the ratio between the (dimensionless) shear and bulk viscosities. Using
Eqs.~(\ref{xi}) and (\ref{eta}), $R$ can also be explicitly written as
\begin{equation}
R = \frac{{\tilde \eta}_0}{{\tilde \xi}_0} \left(\frac{\Omega_v}{\Omega_{v0}}\right)^{\lambda-\nu}.
\end{equation}

Note that in the absence of bulk viscosity, $\omega_v\to 0$, $\omega_v R \to -\tilde{\eta} H/(3H_0 \Omega_v)$
and the above expressions for the factors $A$, $B$, $C$ and $D$ reduce to
\begin{eqnarray}
A &=&  \frac{2\tilde{\eta}}{3 \Omega_v^2} \frac{H}{H_0},
\label{Aeta}
\\
\nonumber \\
B &=&   \frac{4 \tilde{\eta}}{27 a^2 \Omega_v H H_0},
\label{Beta}
\\ 
\nonumber \\
C &=& -\frac{3}{2} {\Omega_{v}} \frac{H_0^2}{H^2} ,
\label{Ceta}
  \\ 
\nonumber \\
D &=& 0,
\label{Deta}
\end{eqnarray}
and we can see explicitly how the differential equation for the density contrast
depends on the shear.

\section{Numerical Results and Statistical Analysis}
\label{results}

We now present our numerical
results concerning how the density contrast $\Delta$ is changed as a function of
the bulk and shear viscosities. We analyze the behavior in terms of both the
magnitude for the bulk and shear viscosities, $\xi_0$ and $\eta_0$, but also
how the dependence of the viscosities in terms of the fluid density, parameterized 
by the exponents $\nu$ and $\lambda$ in Eqs.~(\ref{xi}) and (\ref{eta}), influence
the results. {}Finally, we will provide a statistical analysis to determine 
preferable values for the viscosities using the most recent
redshift-space-distortion based $f(z) \sigma_8(z)$ data.

\subsection{Linear growth of viscous dark matter halos}

In the following we show the results for the linear evolution of the density contrast
$\Delta$ considering the scale $k=0.2h \,Mpc^{-1}$, which corresponds to the scale for
typical galaxy clusters. According to the $\Lambda$CDM standard cosmology,
such objects became nonlinear, i.e., the density contrast approaches $\Delta \sim 1$
at recent times, when the scale factor is $a_{nl} \sim 0.5$, or equivalently,
at redshift $z_{nl} \sim 1$.
In all of our results the initial conditions are taken at the matter-radiation 
equality and they are set with the help of the CAMB code~\cite{CAMB}.

\begin{center}
\begin{figure}[!htb]
\subfigure[Dimensionless viscosities set at the value of $10^{-6}$.]{\includegraphics[width=7.5cm]{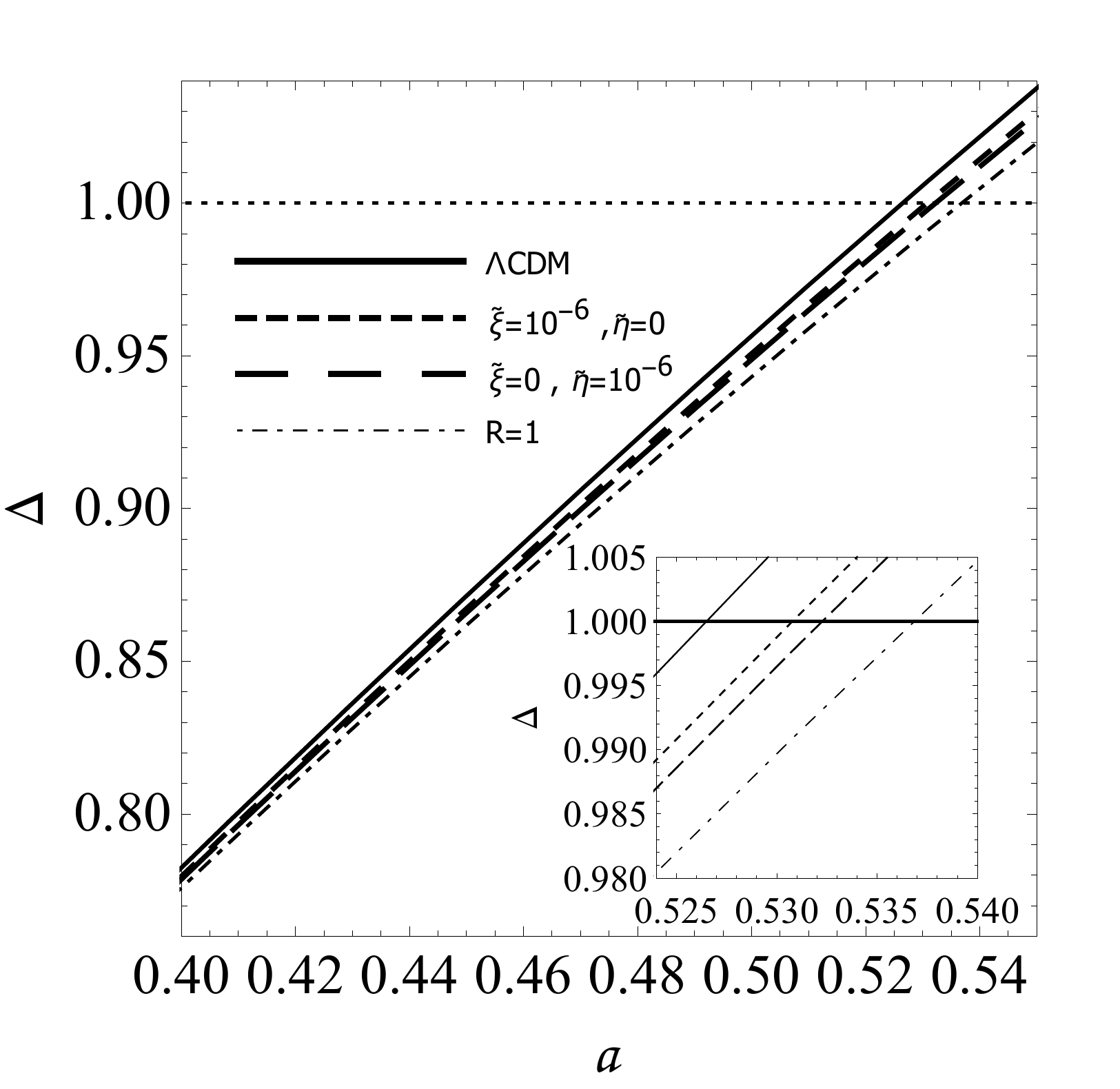}}
\subfigure[Dimensionless viscosities set at the value of $10^{-5}$.]{\includegraphics[width=7.5cm]{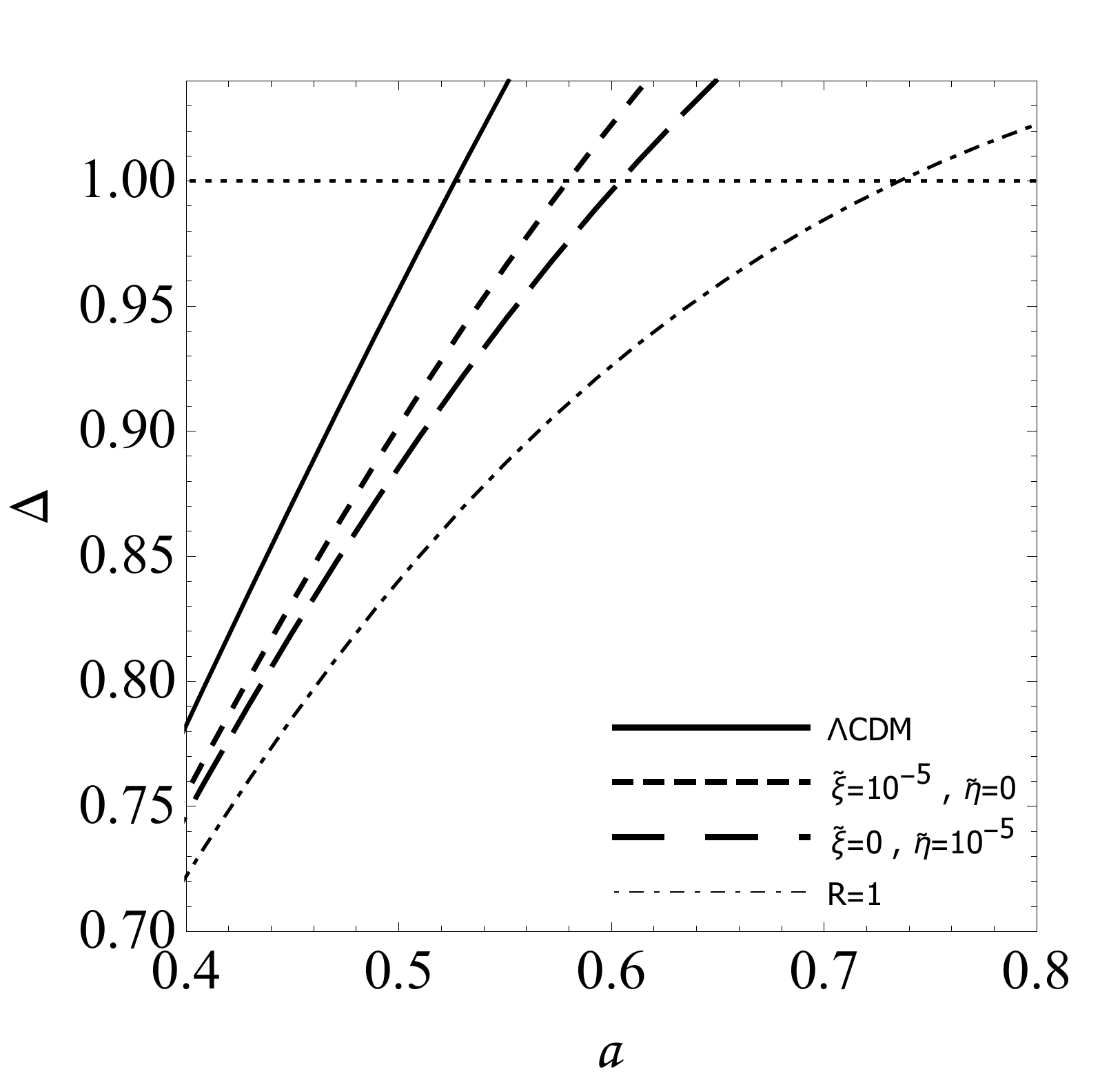}}
\caption{The density contrast $\Delta$ as a function of the scale factor $a$ showing the combined 
effects of bulk and shear for two different values for the dimensionless viscosities 
(when setting one of them to zero) and the comparison with the case $R=1$, i.e., $\tilde{\xi}=\tilde{\eta}$,
and the standard $\Lambda$CDM case. }
\label{k02ab}
\end{figure}
\end{center}

In fig.~\ref{k02ab} we compare the combined effects of bulk and
shear with their isolated effects. Here, for convenience, we have assumed 
constant coefficients, by taking the exponents $\nu=\lambda=0$ in Eqs.~(\ref{xi}) and (\ref{eta}).
We can see clearly from the results shown in this figure that both bulk and shear contribute 
in a similar way. Both viscosities act in a way to attenuate the matter perturbation growth. 
{}For the case $\tilde{\eta}_0=10^{-6}$ and
$\tilde{\xi}_0=10^{-6}$, panel (a), the evolution of $\Delta$ remains close to the concordance
$\Lambda$CDM model. The $R=1$ curve means that both effects are
acting simultaneously with the same magnitude, i.e., $\tilde{\eta}_0=\tilde{\xi}_0=10^{-6}$
in panel (a) and $\tilde{\eta}_0=\tilde{\xi}_0=10^{-5}$
in panel (b).

\begin{center}
\begin{figure}[!tb]
\subfigure[Including only the shear viscosity.]{\includegraphics[width=7.5cm]{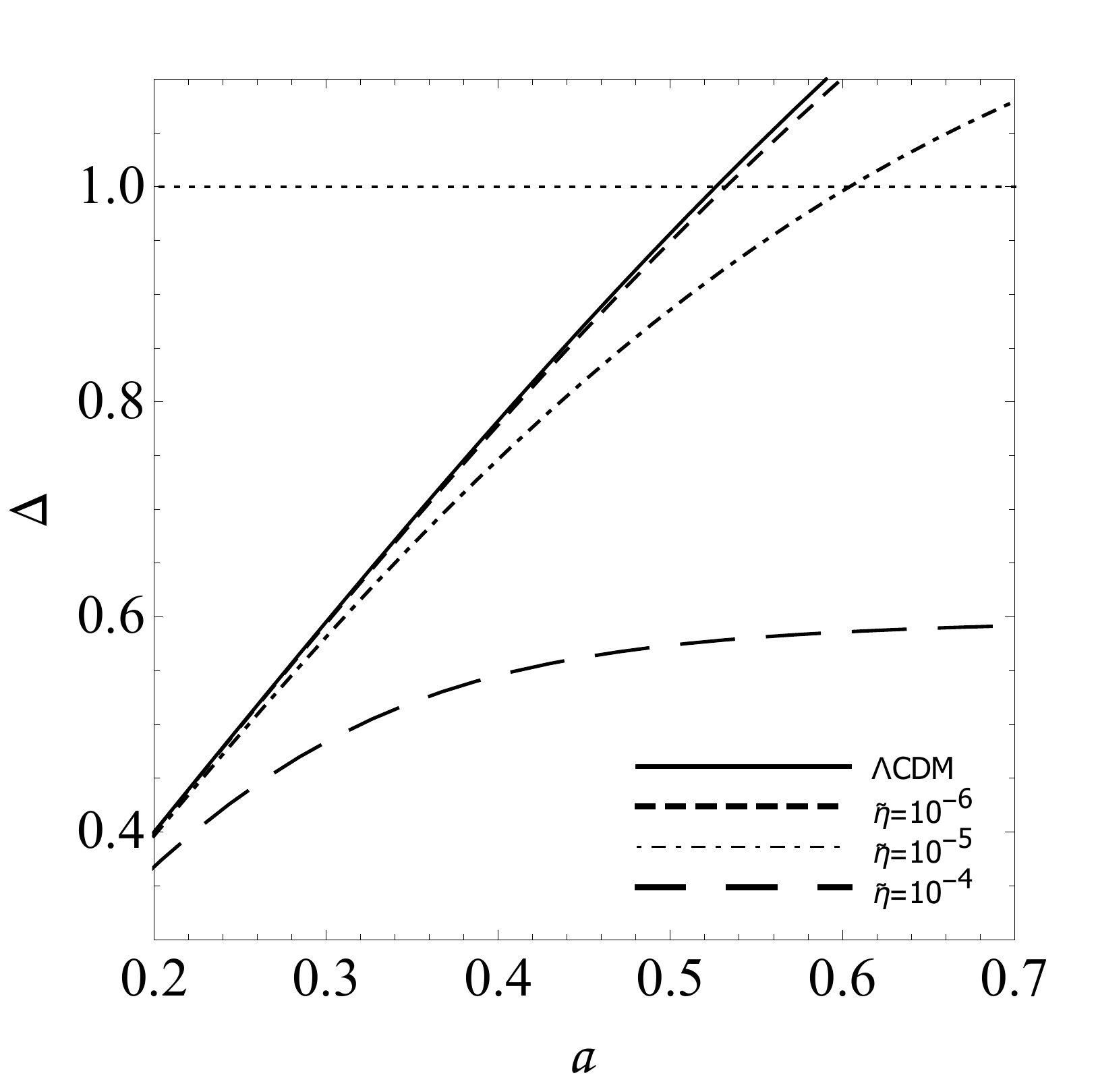}}
\subfigure[Including only the bulk viscosity.]{\includegraphics[width=7.5cm]{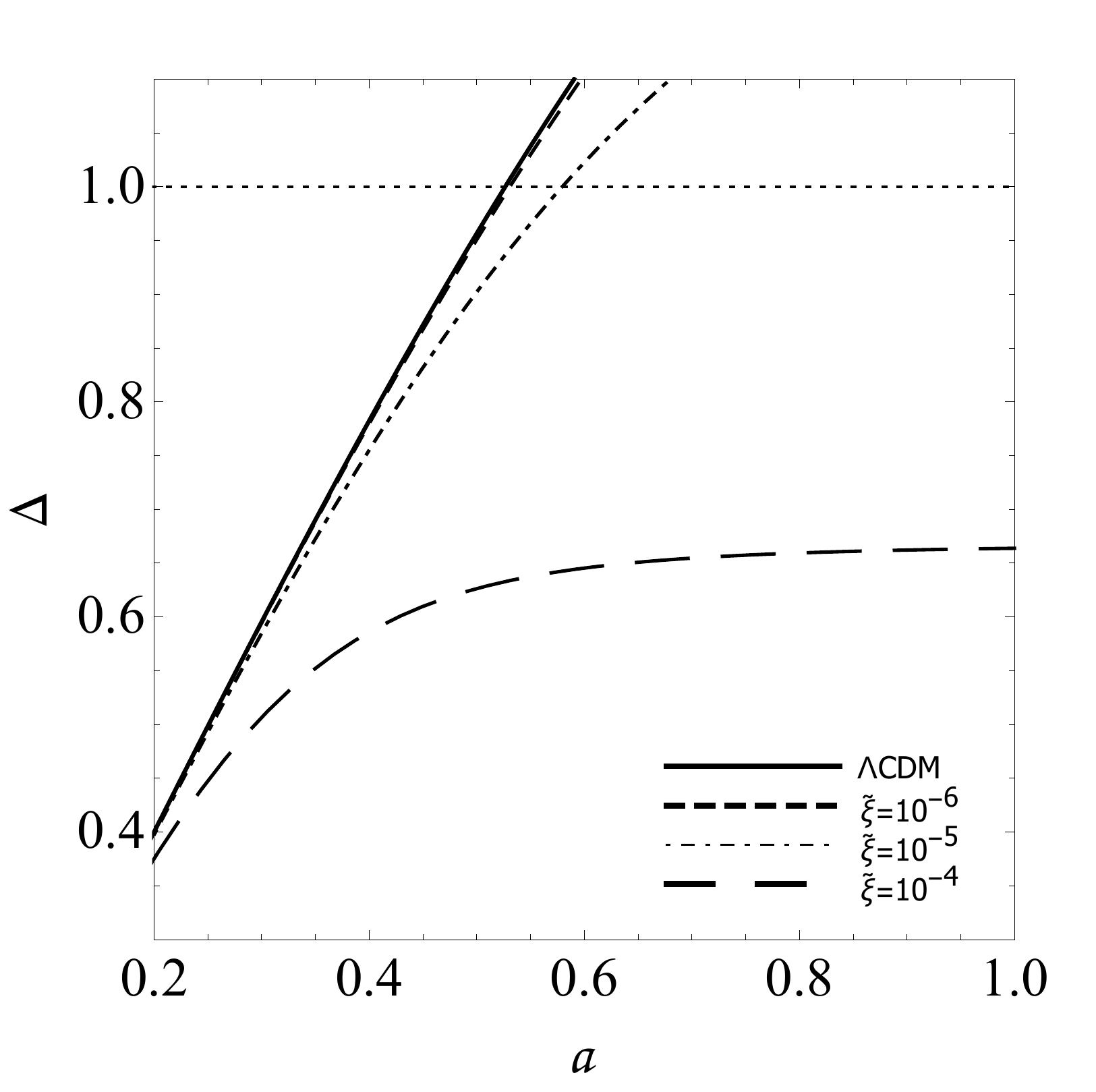}}
\caption{The density contrast $\Delta$ as a function of the scale factor $a$ showing the individual
effects due to bulk and to the shear viscosity. }
\label{k02nulo}
\end{figure}
\end{center}

In fig.~\ref{k02nulo} we show the isolated effects of the shear, shown in the panel (a), and that due to the bulk,
shown in the panel (b), on the density contrast $\Delta$. As in the previous figure, we are also here assuming 
constant viscosities, i.e., we have considered $\nu=\lambda=0$ in Eqs.~(\ref{xi}) and (\ref{eta}). 
It is clear that both the shear and the bulk act in a similar way on how they suppress (damp)
the growth of $\Delta$.

\begin{center}
\begin{figure}[!h]
\subfigure[Vanishing shear viscosity and varying the bulk viscosity
exponent $\nu$ with positive values.]{\includegraphics[width=7.5cm]{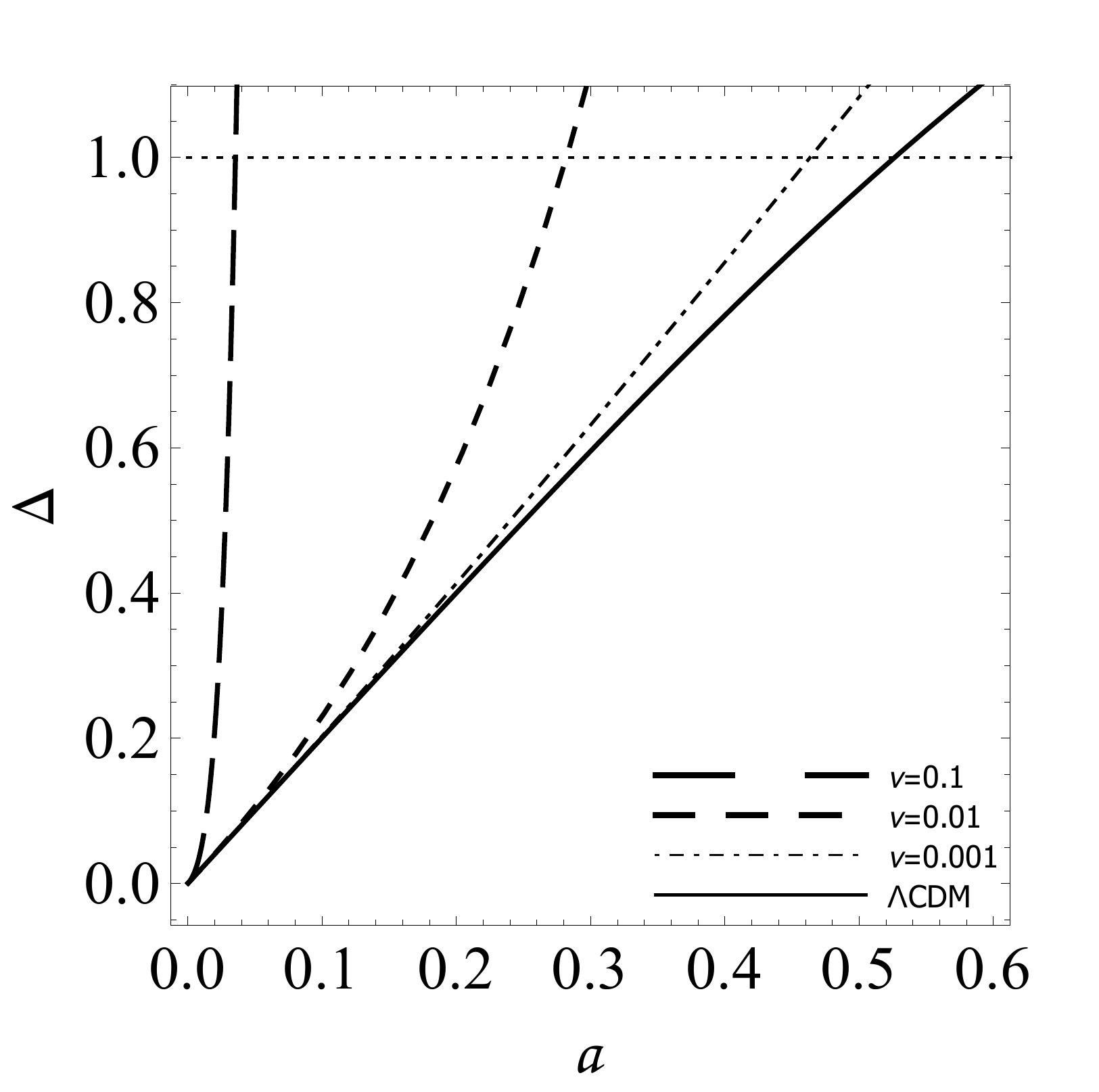}}
\subfigure[Fixed value of $\nu$ and varying ratio $R$.]{\includegraphics[width=7.5cm]{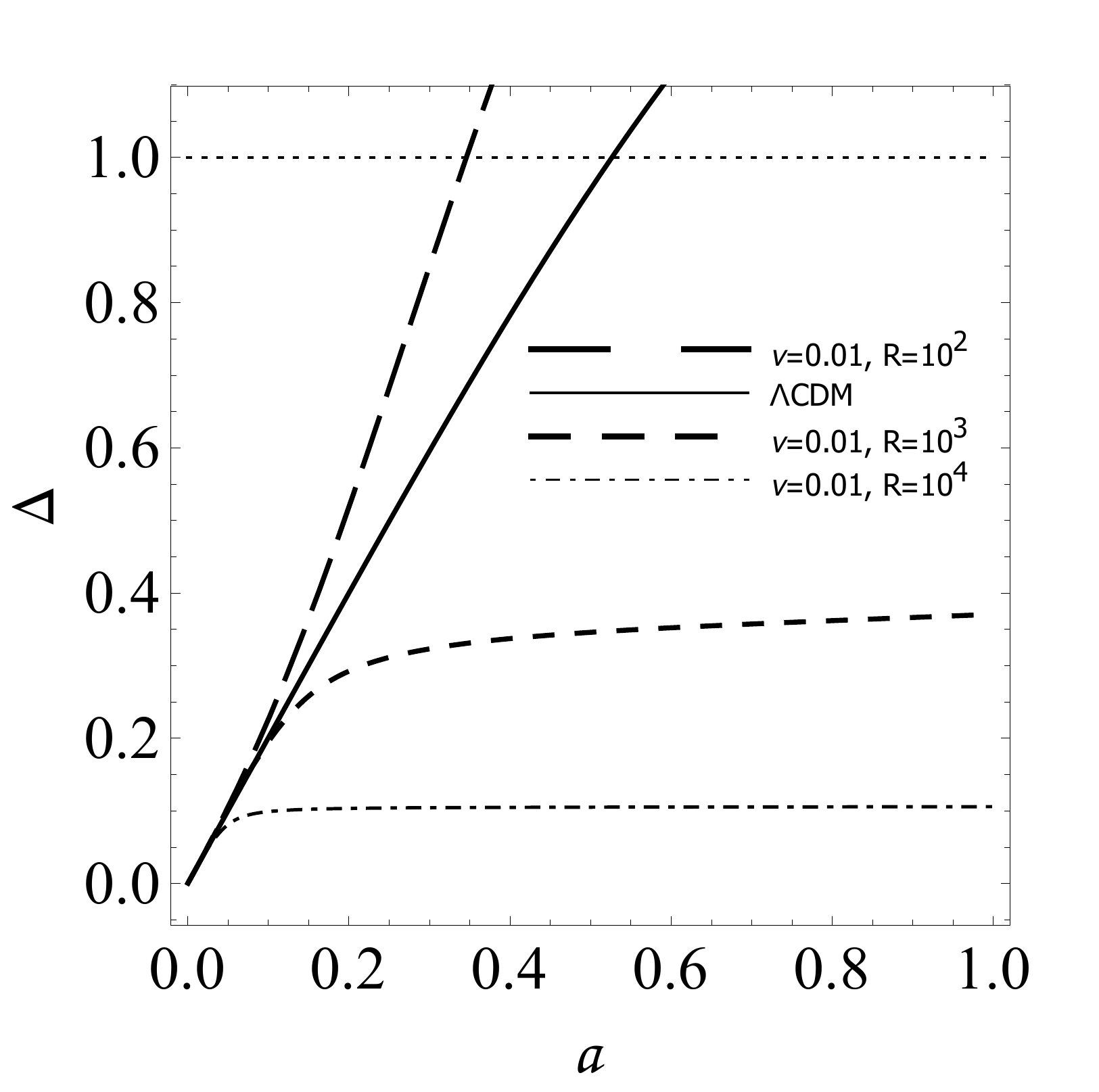}}
\subfigure[Vanishing shear viscosity and varying the bulk viscosity
exponent $\nu$ with negative values.]{\includegraphics[width=7.5cm]{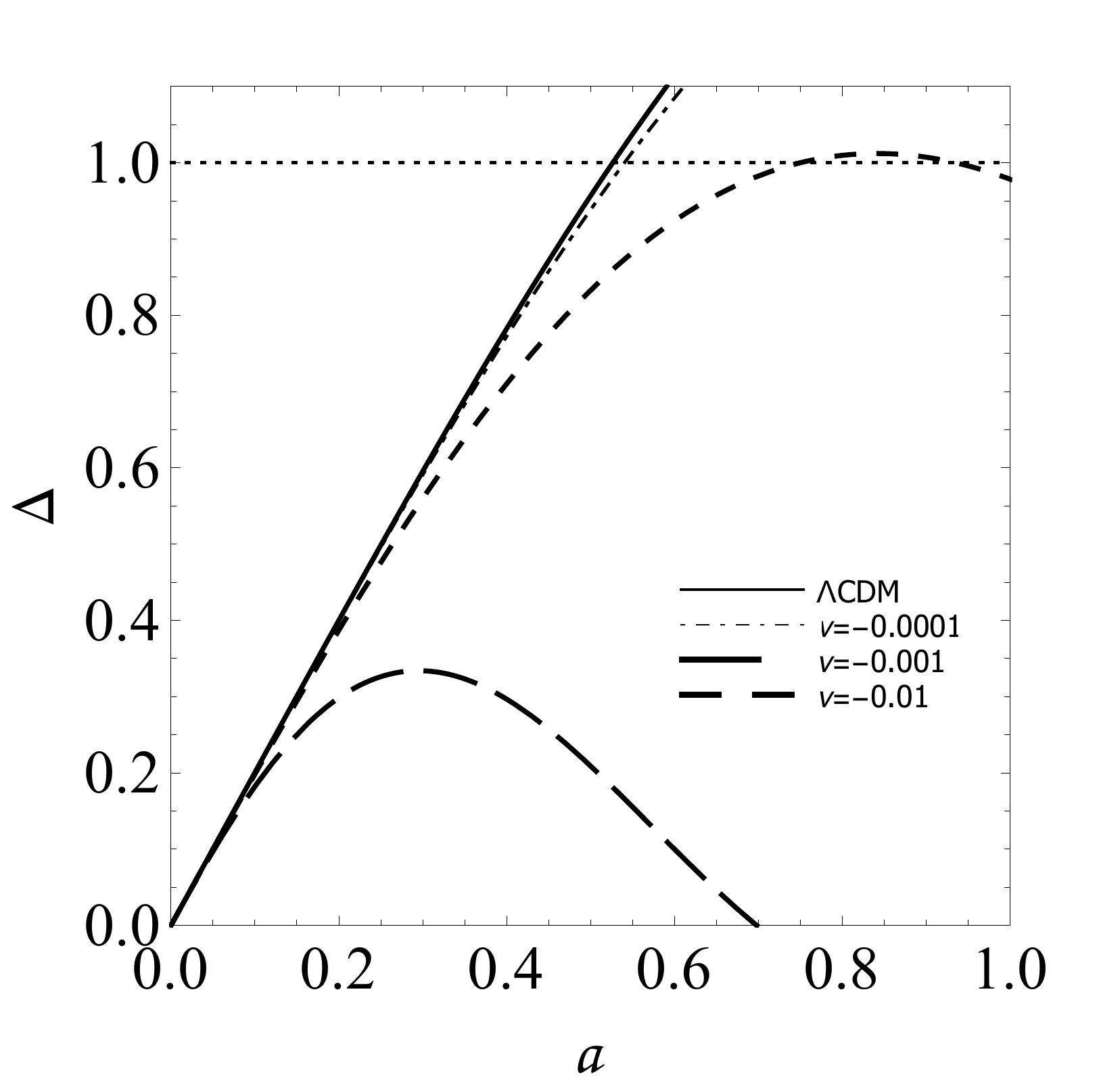}}
\subfigure[Fixed bulk viscosity
exponent $\nu = -10^{-4}$ and varying the ratio $R$.]{\includegraphics[width=7.5cm]{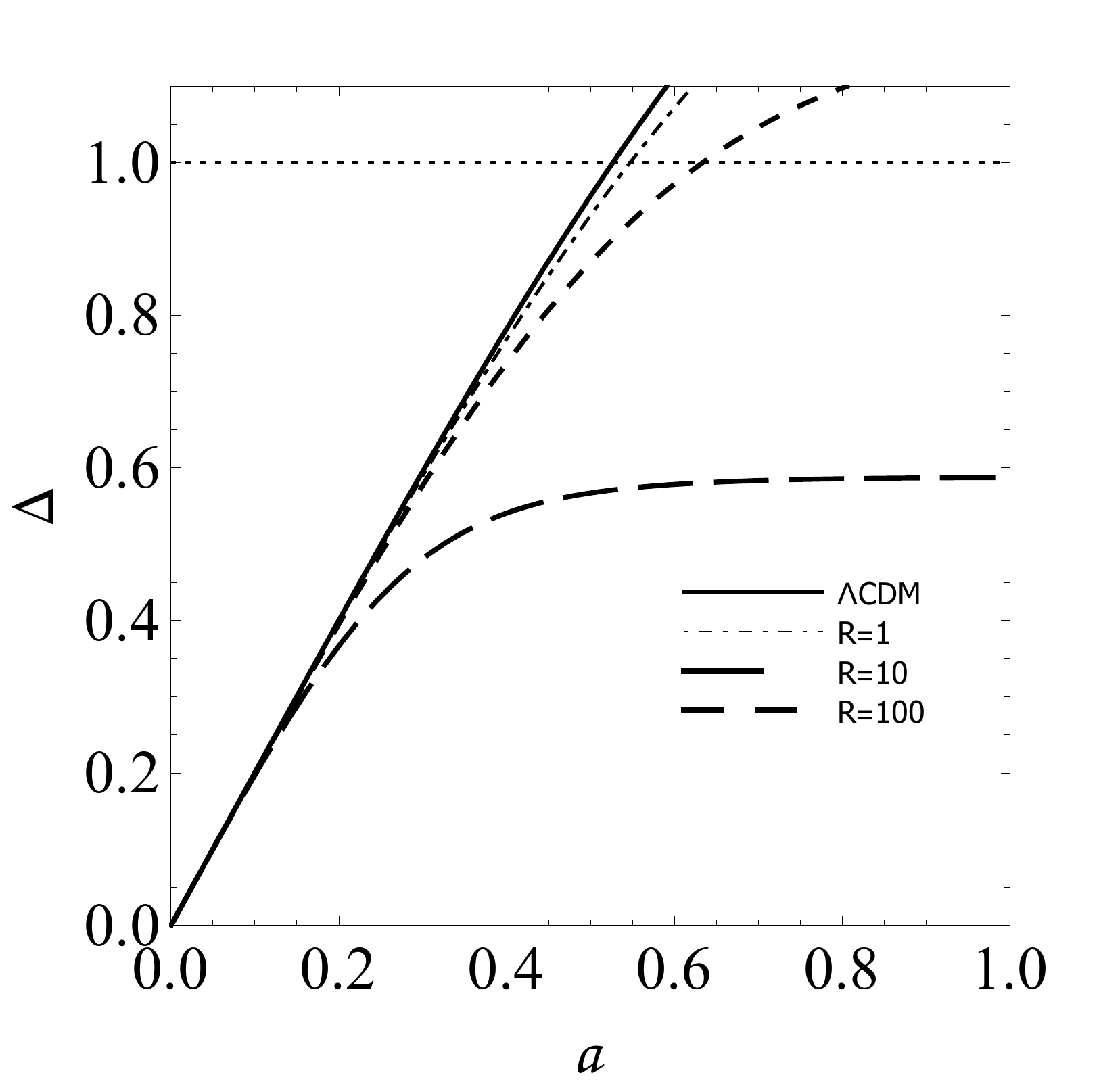}}
\caption{The density contrast $\Delta$ as a function of the scale factor $a$ for fixed bulk viscosity
at the dimensionless value  $\tilde{\xi}_0=10^{-6}$ and for different exponent $\nu$ and ratio $R$.
The shear viscosity exponent in all cases is $\lambda = 0$.}
\label{k2nu}
\end{figure}
\end{center}

In general, bulk and shear viscosities act to suppress the growth of
density perturbations, as it could be expected from the beginning. It
must be stressed once again that even though the shear does not contribute for the
background dynamics, it affects the perturbations. The contribution of the
shear viscosity for the suppression of the growth of the matter perturbation 
is seen as effective as the bulk viscosity is and it might be even more relevant, 
as the results shown in figs.~\ref{k02ab} and \ref{k02nulo} indicate. 
The simultaneous combination of both dissipative
effects enhances the suppression of the matter perturbation growth. {}For the perturbations 
to be able to reach the nonlinear regime
(as required to give birth to local structures such as galaxies and
clusters of galaxies), the dimensionless coefficients $\tilde{\xi}$ and
$\tilde{\eta}$, for the bulk and shear viscosities, respectively, when acting individually or in
combination, must be at most of the order of $10^{-6}$. The relative contribution between the two
dissipative components, $R$, is also relevant. {}For instance, when $R$ is greater than
one the perturbations in general do not reach the nonlinear
regime. This reveals the important r\^ole played by the shear viscous
component.

\begin{center}
\begin{figure}[!tb]
\centerline{\includegraphics[width=7.5cm]{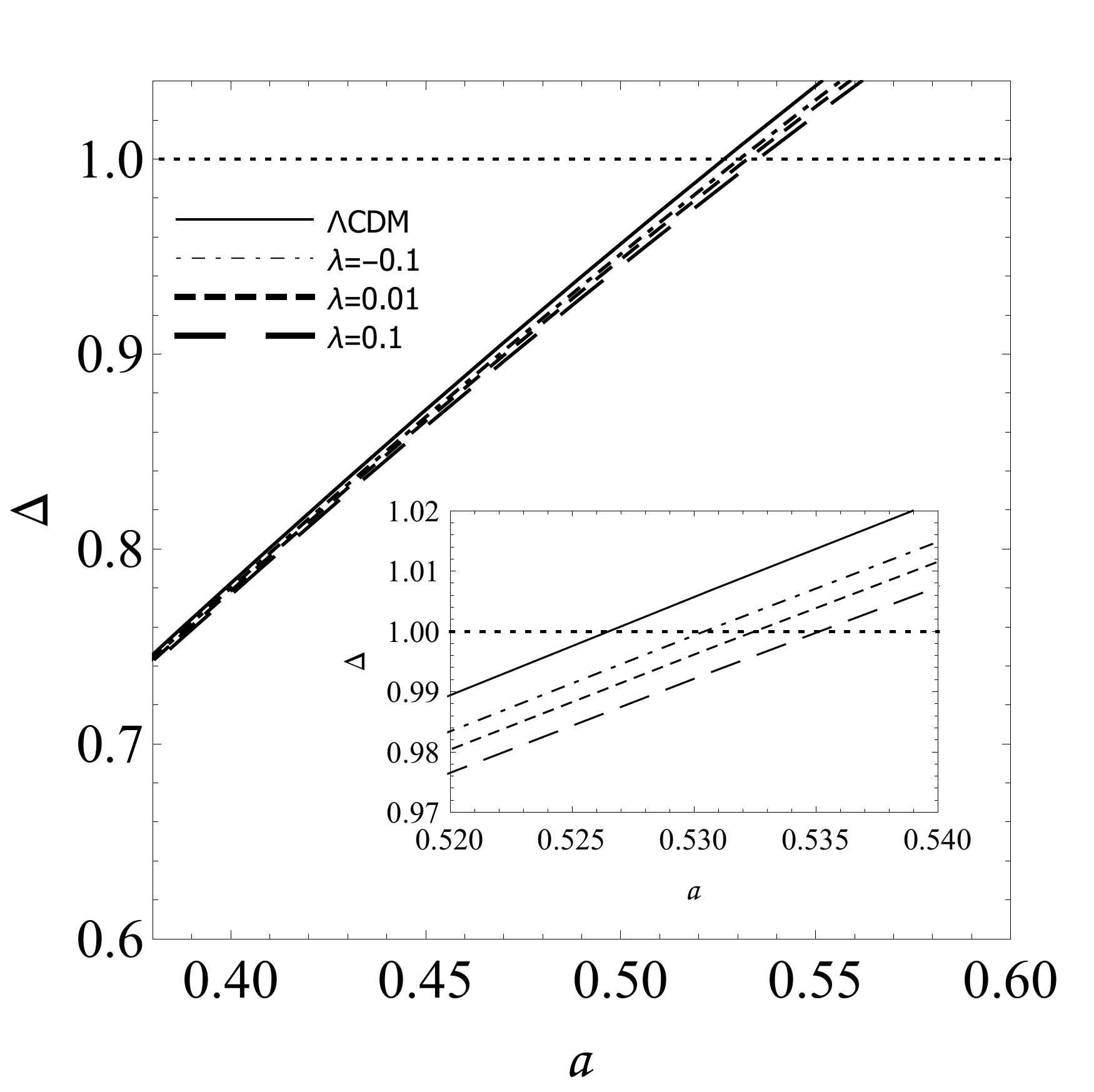}}
\caption{The density contrast $\Delta$ as a function of the scale factor $a$ when the bulk viscosity 
is absent and for different values for the shear viscosity exponent $\lambda$.}
\label{k02lambda}
\end{figure}
\end{center}

In fig.~\ref{k2nu} we analyze the effect of the dependence set for the bulk viscosity in Eq.~(\ref{xi}) through its
exponent $\nu$. Surprisingly, we find that the presence of the bulk viscosity is able to lead to
an enhancement, as opposite to suppression, when the
shear is absent and $\nu$ is positive. This is seen explicitly in the panels (a) and (b) of fig.~\ref{k2nu}.
This anomalous behavior has
already been detected in Ref.~\cite{Velten:2013pra}. Note, however, this same trend does
not happen for the case of the shear viscosity, where it always leads to a suppression of the perturbations
for either positive or negative exponent $\lambda$ in Eq.~(\ref{eta}) and when the bulk is absent.
This is explicitly shown by the results displayed in fig.~\ref{k02lambda}. The inset in fig.~\ref{k02lambda}
zoom in the region around the transition to nonlinearity for the perturbations. Note, however,
that these results also show that a shear viscosity
with a negative exponent $\lambda$ suppresses less the growth of $\Delta$ than the cases
with a positive $\lambda$, where we see a relatively larger damping of $\Delta$.
This indicates the crucial role played by the shear in the suppression of
power in the matter agglomeration in general situations.

\begin{center}
\begin{figure}[!htb]
\subfigure[Varying the bulk viscosity
exponent $\nu$ at vanishing shear viscosity.]{\includegraphics[width=7.5cm]{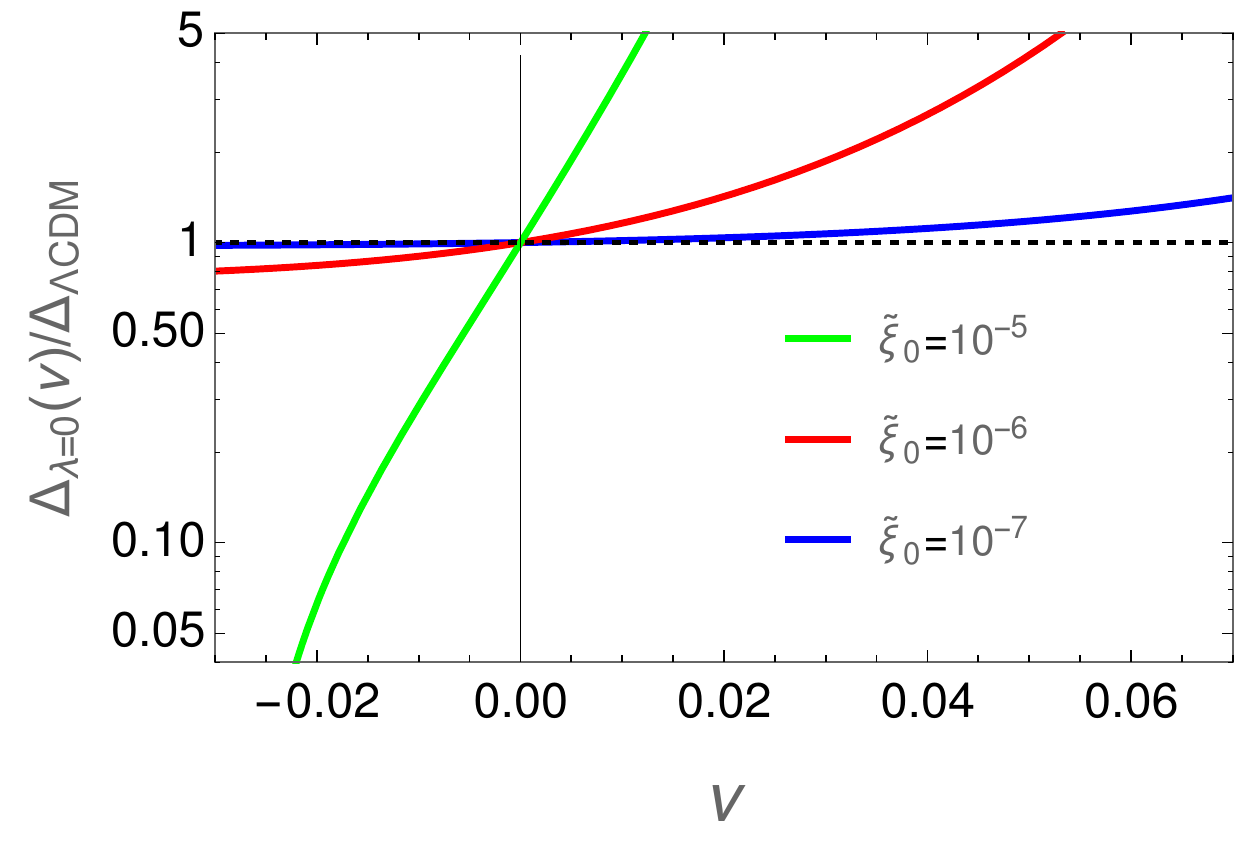}}
\subfigure[Varying the shear viscosity
exponent $\lambda$ at vanishing bulk viscosity.]{\includegraphics[width=7.5cm]{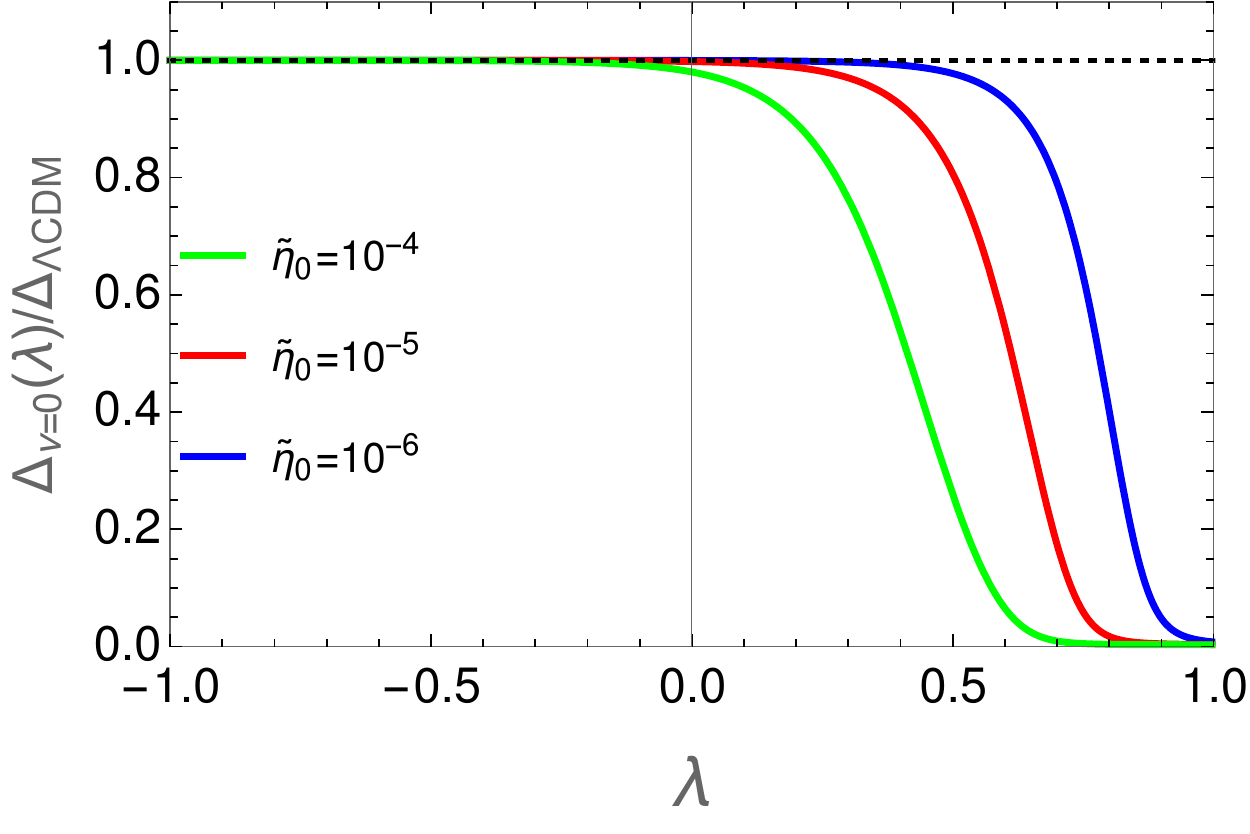}}
\caption{The ratio of the density contrast $\Delta$ with respect the $\Lambda$CDM value, both fixed at the
scale $a=0.1$, as a function of the viscosities exponents $\nu$ and $\lambda$ and for different values 
for the viscosity coefficients. }
\label{nulambda}
\end{figure}
\end{center}

Both of these results observed with the bulk and the shear with respect to the sign of the exponents 
in Eqs.~(\ref{xi}) and (\ref{eta}) can be understood from the equations derived in the previous 
section. 
The bulk viscosity influences both the damping term (the first-order derivative term) in Eq.~(\ref{euler})
as also the term in front of the linear term, i.e., the "frequency" term, in an analogy with the damped harmonic
oscillator equation in basic mechanics. The most relevant change  caused by the exponent $\nu$ happens
in particular in the later. The term $C+k^2 D$ is dominated by the last contribution in Eq.~(\ref{termD}).
{}For a bulk viscosity with a {\it positive} exponent $\nu$ it leads to a larger {\it negative}
frequency like-term. Thus, favoring a larger growing mode with respect to $\Lambda$CDM. The opposite happens
with a {\it negative} $\nu$, which increases the frequency-like term positively and causes a larger suppression
effect on the growth of $\Delta$.
This change of behavior for $\Delta$ for a negative or positive $\nu$ explains the results seen in fig.~\ref{k2nu}.
When the bulk viscosity is absent and only the shear viscosity is present, the frequency-like term is unchanged
with respect to the $\Lambda$CDM value and the shear viscosity will affect only the damping-like term
in Eq.~(\ref{euler}).
When the bulk viscosity is absent, it follows the Eqs.~(\ref{Aeta}), (\ref{Beta}), (\ref{Ceta}) and (\ref{Deta}).  
Recalling that in this case that the solution of Eq.~(\ref{dOmegav}), when $\omega_v=0$, is
$\Omega_v \propto 1/a^3$. Using Eq.~(\ref{eta}), we can easily observe that for a {\it positive} exponent $\lambda$
the term $A+k^2 B$ in the damping-like term in the differential equation~(\ref{euler}) 
will grow slower with the scale factor than in the case when the
exponent $\lambda$ of the shear
viscosity is {\it negative}. Thus, we have (recalling that for our initial conditions
$a<1$) that the damping-like term for $\lambda>0$ will get larger than in the case when $\lambda<0$. Hence, 
there is a larger damping of $\Delta$ when the
exponent $\lambda$ is positive than in the case when $\lambda$ is negative. Thus, a negative $\lambda$ will always lead
to results that are closer to the $\Lambda$CDM case.  This explains the behavior seen in 
fig.~\ref{k02lambda}.
Both of these behaviors with respect to the exponents $\nu$ and $\lambda$ for the viscosities get evident
also when we present results for the density contrast $\Delta$ at some fixed value of the scale $a$. 
In fig.~\ref{nulambda} we show the ratio of the density contrast $\Delta$ with respect to the $\Lambda$CDM value,
$\Delta_{\Lambda{\rm CDM}}$,
i.e., in the absence of viscosities. {}For convenience, we have fixed the scale at the value $a=0.1$,
for which  $\Delta_{\Lambda{\rm CDM}}\simeq 0.2$ and, thus, well before the nonlinear regime.
The behavior explained above gets evident when analyzing the results in this figure. {}For example, in the panel (a) of
fig.~\ref{nulambda} we see that a negative $\nu$  leads to a result for $\Delta$ that is smaller than in
the $\Lambda$CDM case, while the opposite is seen when $\nu$ is positive, in which case $\Delta > \Delta_{\Lambda{\rm CDM}}$.
Now, in the absence of a bulk viscosity but for a nonvanishing shear viscosity, we have the results shown
in the panel (b) of the same figure. In fig.~\ref{nulambda}(b) we see that a shear viscosity always damp
the density contrast and a positive $\lambda$ produces results that are more strongly damped than for a negative
$\lambda$, which tend to remain closer to the $\Lambda$CDM value.

\begin{center}
\begin{figure}[!htb]
\centerline{\includegraphics[width=7.5cm]{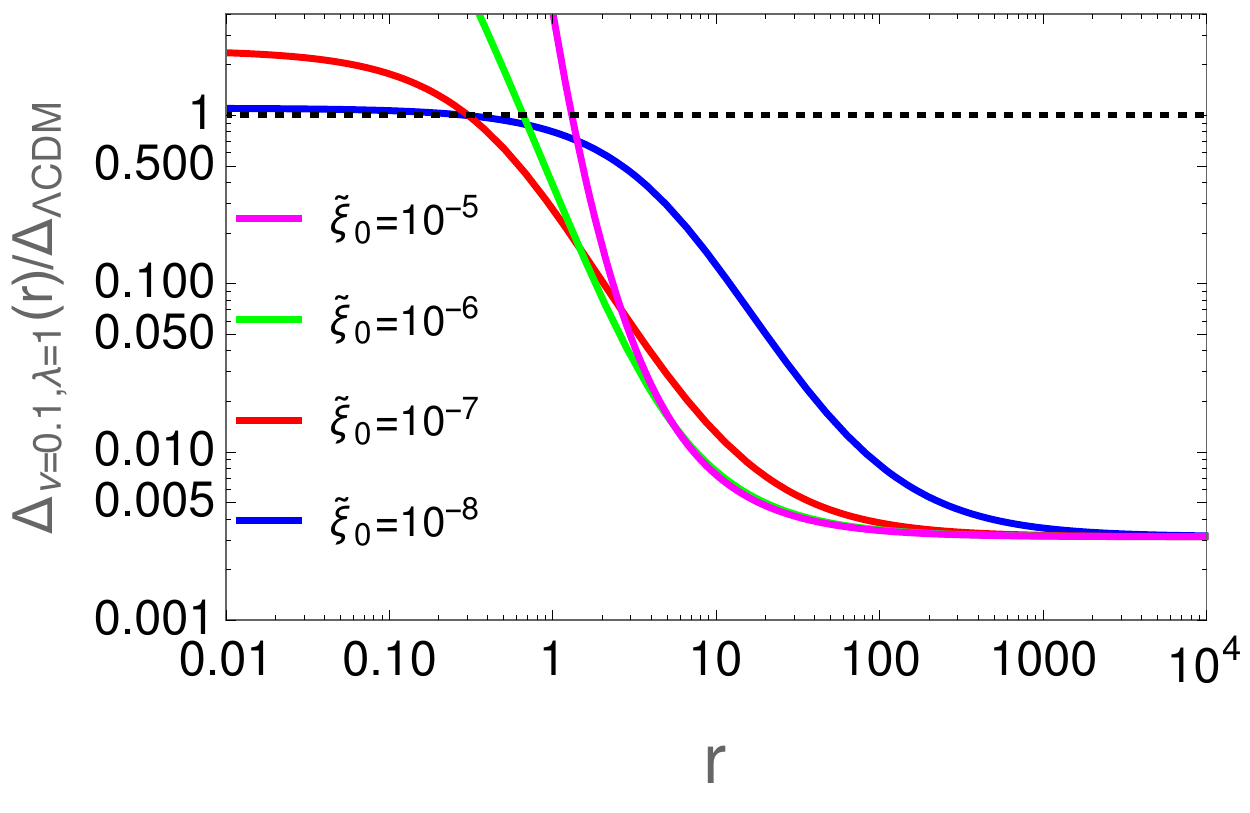}}
\caption{The ratio of the density contrast $\Delta$ with respect the $\Lambda$CDM value, both fixed at the
scale $a=0.1$, as a function of the shear to bulk viscosity (dimensionless) 
coefficients ratio, $r={\tilde \eta}_0/{\tilde \xi}_0$ and for different values 
for the bulk viscosity coefficient. The viscosities exponents have been fixed at the values $\nu=0.1$ and
$\lambda=1$.}
\label{deltar}
\end{figure}
\end{center}

Based on the above results, it is interesting to investigate when a larger viscosity gets favored.
{}From the results shown in fig.~\ref{nulambda} we see that a bulk viscosity 
 with a negative exponent $\nu$ always damp the growth of $\Delta$
and the nonlinear regime is delayed or even prevented for a sufficiently large (and negative) $\nu$ and some given
bulk viscosity amplitude $\xi_0$. 
On the other hand, a bulk viscosity with a positive exponent $\nu$ leads to a nonlinear regime for $\Delta$ 
that can happen much earlier in the evolution than
one would desire, thus exacerbating the problems that suffers the
$\Lambda$CDM model and explained in the Introduction. A positive exponent $\nu$ for the bulk viscosity
seems, thus, disfavored.
As for the shear viscosity, we always has a suppression of the density contrast with respect to the 
$\Lambda$CDM case. {}Furthermore, a positive and larger exponent $\lambda$ damps $\Delta$ more
strongly than a negative $\lambda$. 
Hence, we see that the larger growth of the density contrast observed with a bulk viscosity
with a positive $\nu$ can be compensated with a shear viscosity that also has a positive
exponent $\lambda$, which is the case that leads to the larger damping of $\Delta$.
This state of affairs caused by the combined effects of both bulk and shear viscosities
is exemplified in the results shown in fig.~\ref{deltar}, where we show the same ratio
of density contrast used in fig.~\ref{nulambda} but now given as a function of the magnitude
of the shear viscosity, $\eta_0$. The results are presented in terms of the ratio $r$, where
$r= {\tilde \eta}_0/{\tilde \xi}_0$, i.e., the ratio between the (dimensionless) magnitude for
the shear and bulk viscosities. We have once again fixed the scale at the value $a=0.1$ 
for convenience and we have chosen the bulk and shear viscosities coefficients at the values $\nu=0.1$ and $\lambda=1$,
respectively.
{}From the results shown in fig.~\ref{deltar} we see that we can easily compensate the growing behavior
of $\Delta$ due to a bulk viscosity with a positive and large exponent $\nu$ with an equally large and positive 
shear viscosity exponent $\lambda$ and a larger magnitude for the shear viscosity.
This is mostly likely a more natural situation from a physical view point, since viscosities tend
to increase with the density (see, e.g., Refs.~\cite{Weinberg:1972,Weinberg:1971}) and not the opposite
(i.e., viscosities that decrease when the fluid density increases, as in the case where the
exponents $\nu$ and $\lambda$ are negative).   

\subsection{Constraints from redshift space distortions}

The features presented by dissipative effects on the linear
perturbation theory can also be studied via the growth rate of matter
fluctuations data. Observational projects have inferred from large
scale clustering  the redshift-space-distortion based
$f(z) \sigma_8(z) $ at different
redshifts. This observable combines the linear growth rate $f$,

\begin{equation}
D(a) =
\frac{\Delta(a)}{\Delta(a_0)} \hspace{1cm}\Rightarrow \hspace{1cm}
f(a) \equiv \frac{d\ln{D(a)}}{d\ln{a}},
\end{equation}
with the variance $\sigma^2$ of the density field smoothed on $8
h^{-1}$ Mpc scales. The value of the scale factor today is taken as 
$a_0=1$. The sample for the redshift-space-distortion based
$f(z) \sigma_8(z) $ that we consider has 21 data and the values are 
shown in table~\ref{data}. 

We have inspected in the previous subsection the evolution of the mode $k = 0.2 h$Mpc$^{-1}$, which is in the borderline 
dividing the linear and the nonlinear regime today. The results obtained are consistent with the 
initial hypothesis of the validity of the linear regime. Hence, in order to compare with the redshift 
distortion observational data, which refers to modes that stay in the linear regime up to $z \sim 0$, 
it is more convenient to now consider the value of $k = 0.1$Mpc$^{-1}$, which is quite far 
from the nonlinear regime until today and all the expressions derived in the previous section remain
valid.

\begin{table}
\caption{The redshift-space-distortion data considered in our statistical analysis.}
\begin{center}
  \begin{tabular}{ | c | c | c | }
    \hline
    $z$ & $f(z)\sigma_8(z)$ & Reference \\ \hline
    $0.02$ & $0.360 \pm 0.040$ & \cite{Hudson:2012gt} \\ \hline
    $0.067$ & $0.423\pm 0.055$ & \cite{Beutler:2012px} \\ \hline
    $0.10$ & $0.37\pm 0.13$ & \cite{Feix:2015dla} \\ \hline
    $0.17$ & $0.51\pm 0.06$ & \cite{Song:2008qt} \\ \hline
    $0.22$ & $0.42\pm 0.07$ & \cite{Blake:2011rj} \\ \hline
    $0.25$ & $0.3512\pm 0.0583$ & \cite{Samushia:2011cs} \\ \hline
    $0.30$ & $0.407\pm 0.055$ & \cite{Tojeiro:2012rp} \\ \hline
    $0.32$ & $0.427\pm 0.056$ & \cite{Gil-Marin:2016wya} \\ \hline
    $0.35$ & $0.440\pm 0.050$ &  \cite{Song:2008qt} \\ \hline
    $0.37$ & $0.4602\pm 0.0378$ &   \cite{Samushia:2011cs} \\ \hline
    $0.40$ & $0.419\pm 0.041$ &   \cite{Tojeiro:2012rp} \\ \hline
    $0.41$ & $0.45\pm 0.04$ &   \cite{Blake:2011rj} \\ \hline
    $0.50$ & $0.427\pm 0.043$ & \cite{Tojeiro:2012rp} \\ \hline
    $0.57$ & $0.427\pm 0.066$ &   \cite{Reid:2012sw} \\ \hline
    $0.57$ & $0.426\pm 0.029$ &  \cite{Gil-Marin:2016wya} \\ \hline
    $0.6$ & $0.43\pm 0.04$ &   \cite{Blake:2011rj} \\ \hline
    $0.6$ & $0.433\pm 0.067$ &   \cite{Tojeiro:2012rp} \\ \hline
    $0.727$ & $0.296\pm 0.078$ &  \cite{Hawken:2016qcy} \\ \hline
    $0.77$ & $0.490\pm 0.180$ &   \cite{Song:2008qt} \\ \hline
    $0.78$ & $0.38\pm 0.04$ &   \cite{Blake:2011rj} \\ \hline
    $0.80$ & $0.47\pm 0.08$ &   \cite{delaTorre:2013rpa} \\ 
    \hline
  \end{tabular}\label{data}
\end{center}
\end{table}

\begin{center}
\begin{figure}[!htb]
\subfigure[Results when the bulk viscosity is absent.]{\includegraphics[width=7.5cm]{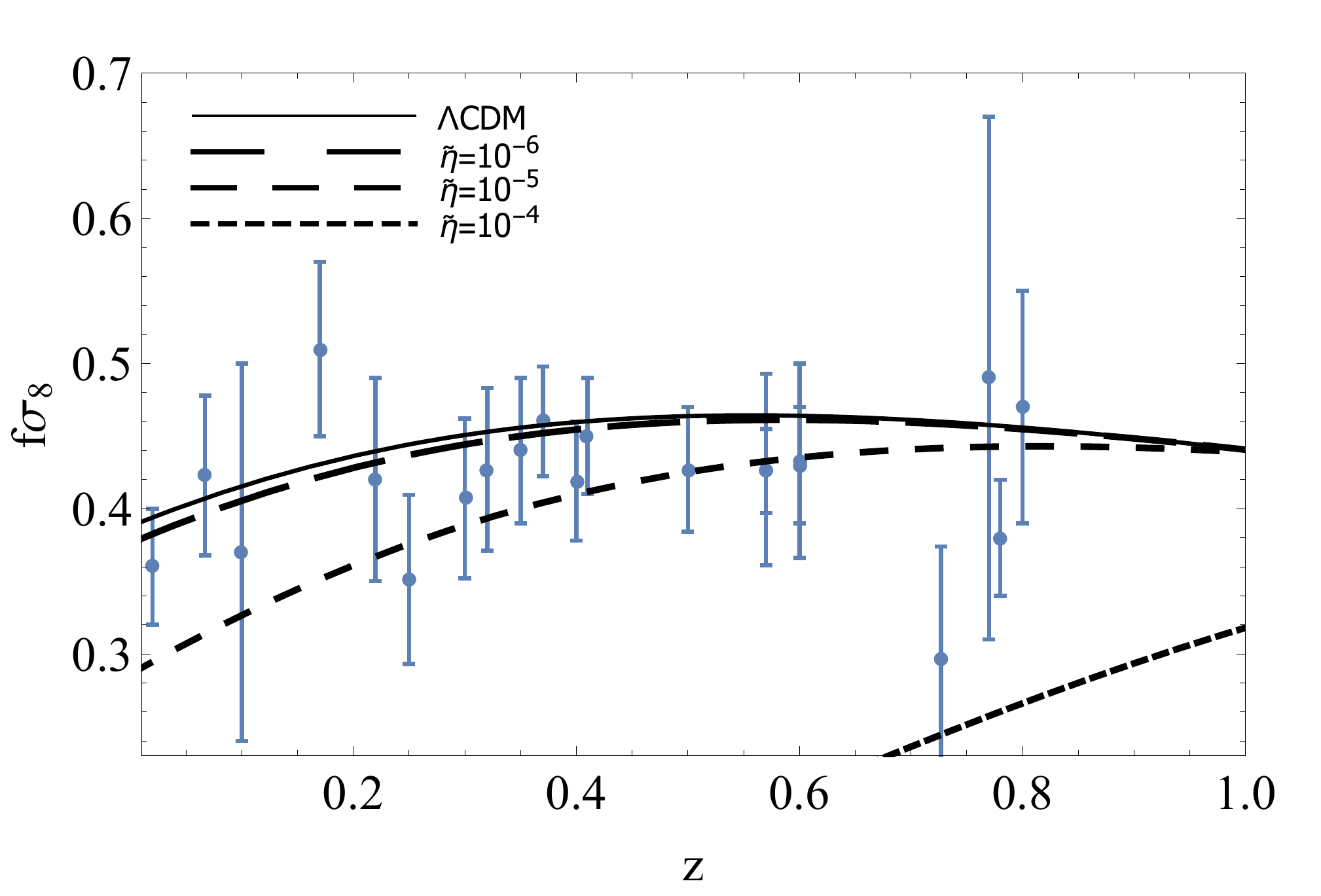}}
\subfigure[Results when the shear viscosity is absent.]{\includegraphics[width=7.5cm]{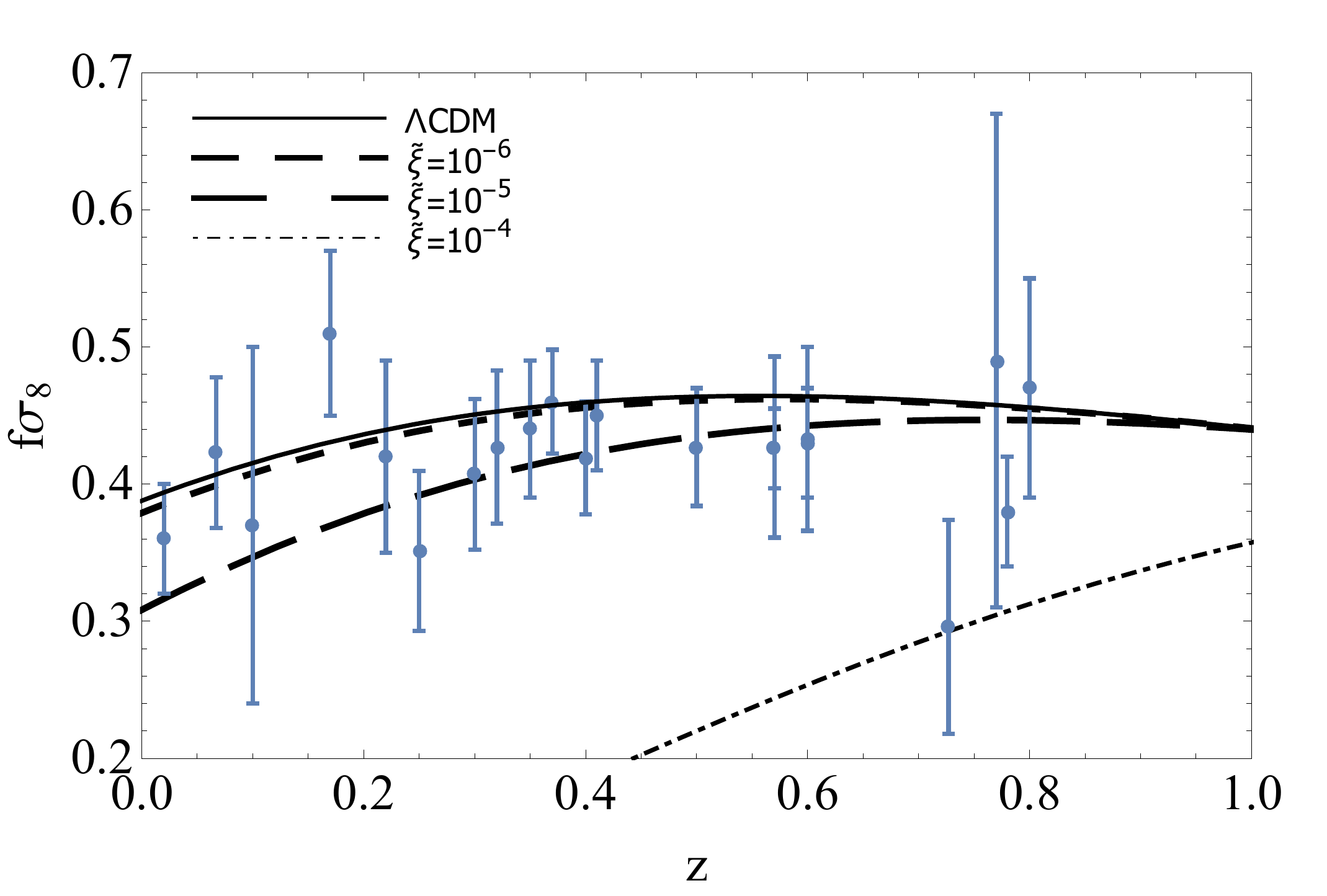}}
\caption{The linear growth against
the $f \sigma_8$ data as a function of the redshift in the absence and presence of the viscosities.}
\label{k01nulo}
\end{figure}
\end{center}

\begin{figure}[!htb]
\centering
\includegraphics[width=7.5cm]{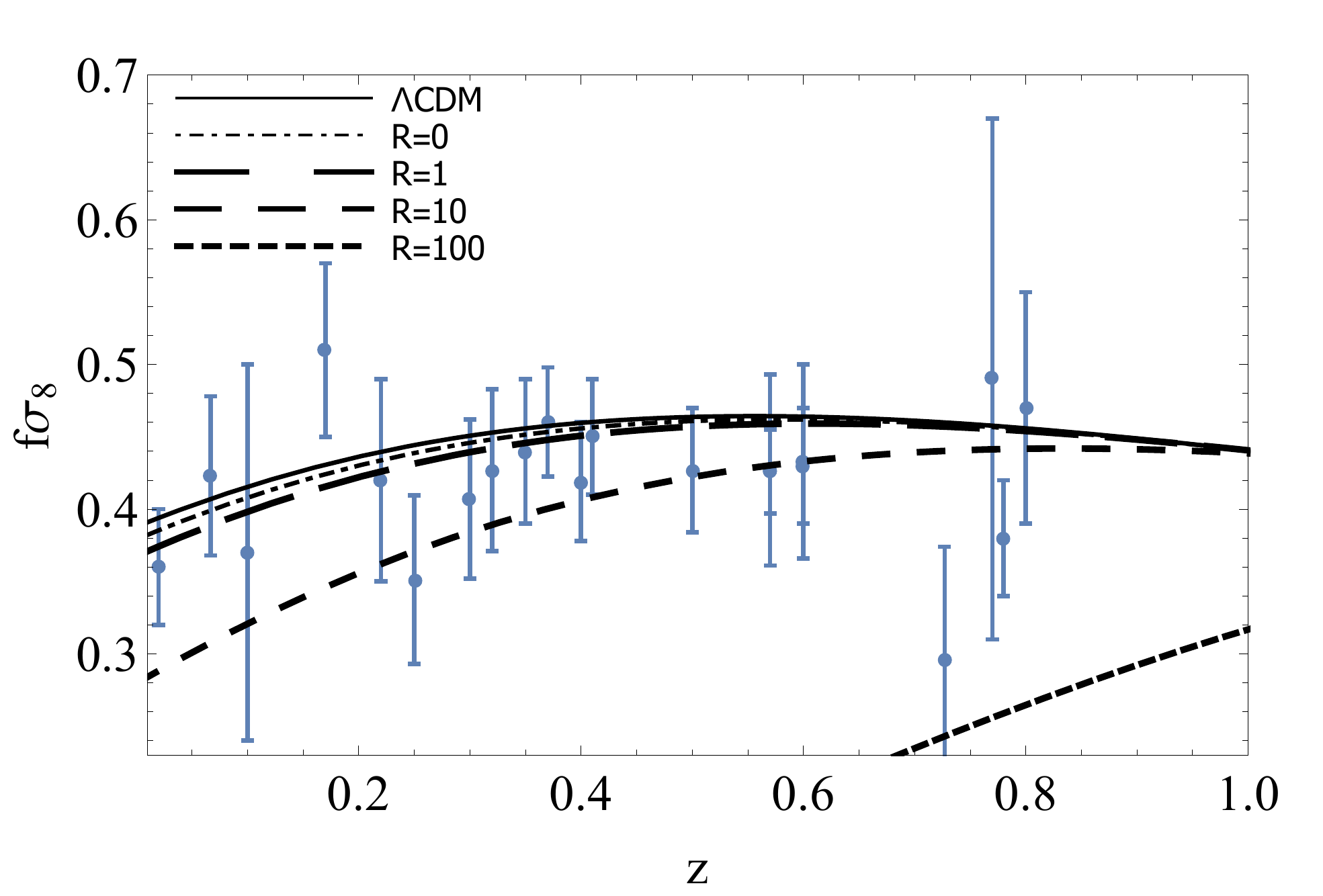}
\caption{The linear growth against
the $f \sigma_8$ data as a function of the redshift with the combined effects of both bulk and shear viscosities.}
\label{k1bulkshear}
\end{figure}

\begin{figure}[!htb]
\centering
\includegraphics[width=7.5cm]{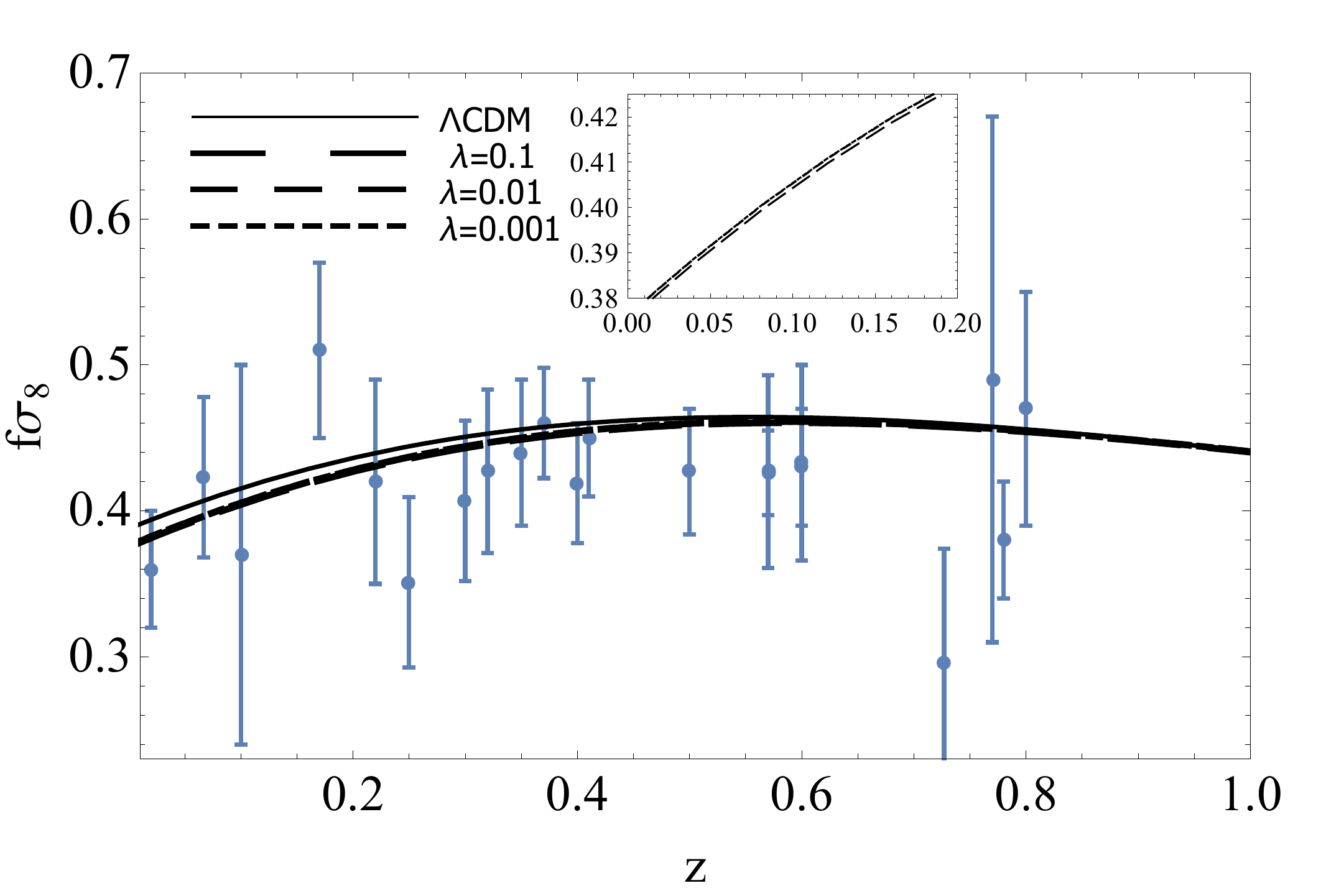}
\caption{The linear growth against
the $f \sigma_8$ data as a function of the redshift for fixed bulk viscosity $\tilde{\xi}_0=0$ and shear $\tilde{\eta}_0=10^{-6}$,
for different values for the shear exponent $\lambda$.}
\label{k1lambda}
\end{figure}

\begin{center}
\begin{figure}[!htb]
\subfigure[Results when the shear viscosity is absent.]{\includegraphics[width=7.5cm]{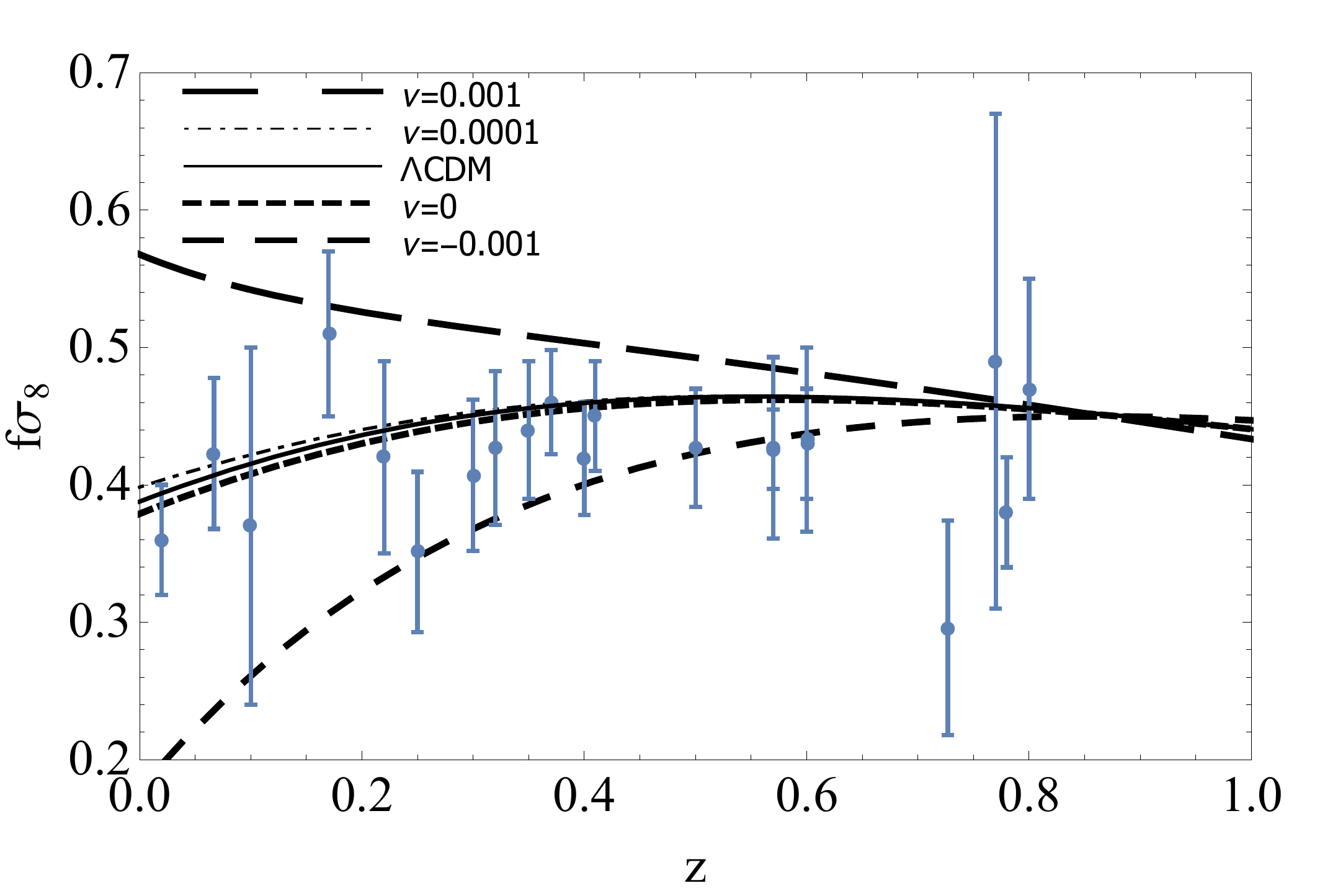}}
\subfigure[Results when $R=1$ and for $\nu=0.01$.]{\includegraphics[width=7.5cm]{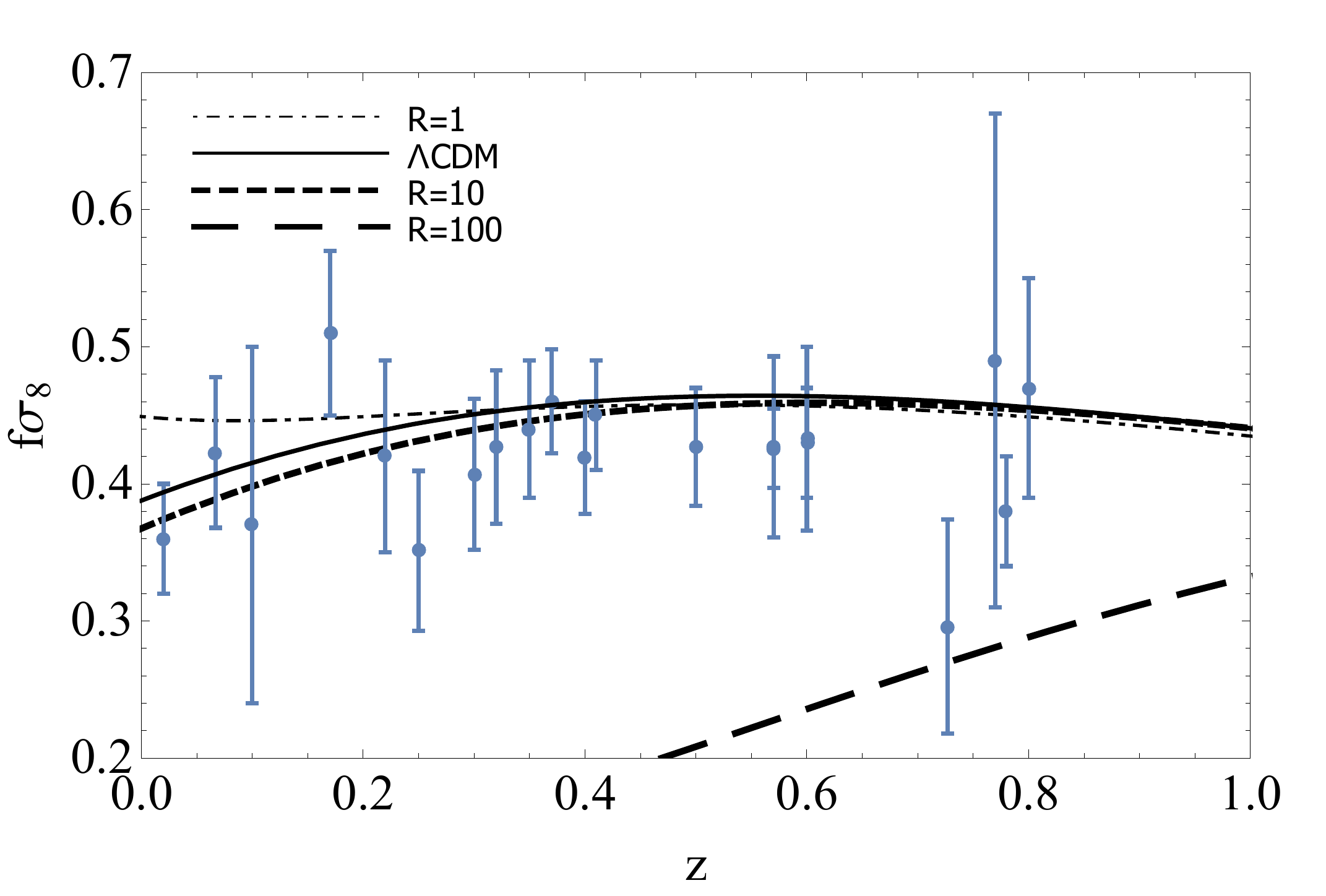}}
\caption{The linear growth against
the $f \sigma_8$ data as a function of the redshift for a fixed value of bulk at $\tilde{\xi}_0=10^{-6}$, but varying $R$ and the
shear exponent $\nu$. }
\label{k01nu}
\end{figure}
\end{center}

The linear growth studied in the previous subsection is now shown against
the $f \sigma_8$ data of table~\ref{data} in figs.~\ref{k01nulo}, \ref{k1bulkshear}, \ref{k1lambda} and \ref{k01nu}. 

\begin{figure}[!htb]
\centering
\includegraphics[width=7.5cm]{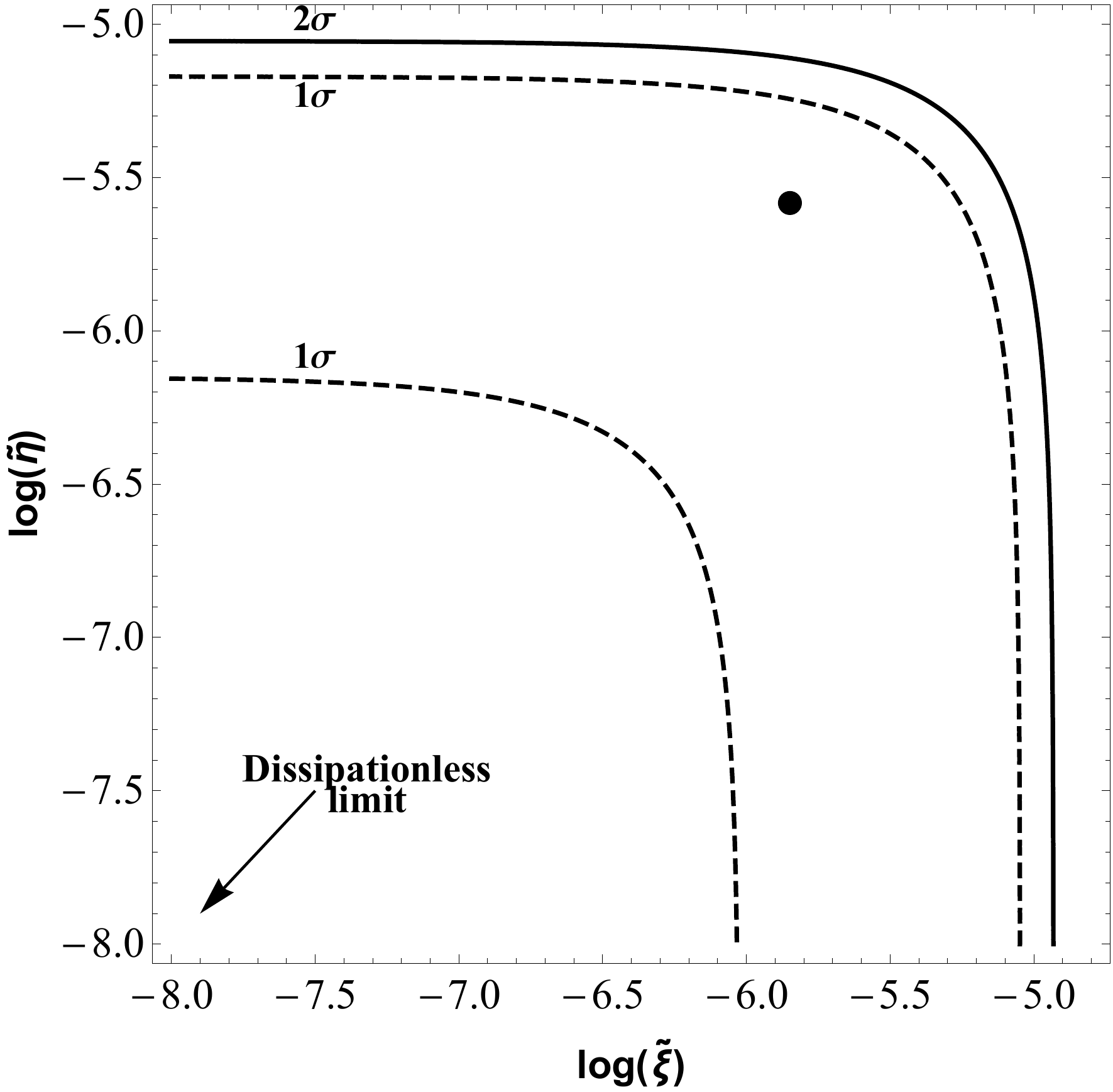}
\caption{ Contours of statistical confidence level at $1\sigma$ and $2\sigma$ using the data from table~\ref{data}. 
The dot indicates the best fit. The standard pressureless case is discarded only within the weak confidence of $1\sigma$. }
\label{contour}
\end{figure}

The confidences levels at
$1\sigma$ and $2\sigma$ are shown in fig.~\ref{contour}, where the best
fitting values for the parameters $\tilde \xi$ and $\tilde \eta$ are also
indicated.
{}For convenience and simplicity of analysis, we consider two
free parameters: The (dimensionless) coefficients for the bulk and the shear viscosities,
$\tilde{\xi}_0$ and $\tilde{\eta}_0$, respectively, 
fixing the respective exponents $\nu$ and $\lambda$ equal to
zero in Eqs.~(\ref{xi}) and (\ref{eta}). Using the $\chi^2$ parameter, 
the best fitting is obtained with
$\chi_{min} = 15.67$, for $\tilde{\xi}_0 = 1.427\times 10^{-6}$ and $\tilde{\eta}_0 =
2.593 \times 10^{-6}$. The striking small value of the $\chi^2_{min}$
per degree of freedom (about $0.55$) is connected with the very large error
bars. The imprecision coming from the redshift space distortions used
here indicates that the observational constraints must be faced with
caution; they have, at the other hand, the advantage of not suffering
too much from model calibration problems, since it is essentially a kinematic effect. 
Even though, we can construct
the parameter estimations using the Bayesian statistical
analysis. Since the values for the coefficients $\tilde \xi$ and $\tilde \eta$ span
many negative order of magnitudes, being very near zero, it is more
convenient to use a logarithmic scale to present the results in fig.~\ref{contour}.  
One notes that the $\Lambda$CDM value ($\xi = \eta = 0$) is
excluded at $1\sigma$, but it is still compatible with the data at $2\sigma$.

\section{The effects of baryons on the dark matter growth}
\label{baryons}

We now investigate whether a separated baryonic component modifies our bounds on the dark matter viscous properties. Indeed, a realistic scenario should take baryons into account. Galaxies are formed when baryons radiate away their kinetic energy falling towards the dark matter potential wells. This process starts already at high redshifts when most of the astrophysical scales of interest are still in the linear regime.
When the nonlinear stage is reached, many distinct processes like for instance radiative pressure feedback, stellar winds, Supernova feedback or local ultraviolet flux from young stars, drive the final evolutionary stage of galaxies. All such physical mechanisms also involve obviously dissipative effects in the baryonic sector, but only a full hydrodynamical simulation 
(including Boltzmann transport equations) can properly investigate the impact of baryons, endowed with such properties, on the final matter clustering patterns and this is beyond the applicability
of linear cosmological perturbation theory used in this work. 
Thus, we stay focused on the linear structure formation regime, where the baryonic sector does not develop such departures from the perfect fluid behavior, i.e., we deal their clustering dynamics as the one of a pressureless fluid.

Including baryons ($p_b=0$) in our model, the Einstein equation $(0-0)$-component is changed, but the $i\neq j$ component is not, since the stress-tensor $T_{ij}$ is not affected by zero pressure components. Thus, the Einstein equations relevant here are,

\begin{equation}\label{Einstein0b}
-k^2 \psi - 3\mathcal{H} \left( \psi^{\prime} +  \mathcal{H} \phi
\right) = \frac{3}{2} \mathcal{H}_0^2 a^2 \left( \frac{\Omega_{b0}}{a^3} \Delta_b +  \Omega_v \Delta_v \right) ,
\end{equation}
and

\begin{equation}\label{Einstein1b}
-\frac{k^2}{2}(\phi - \psi) = \frac{3\mathcal{H}^2}{\rho}\eta \, \theta_v.
\end{equation}
At the background level, it is well established that Big Bang nucleosynthesis (BBN) sets a today's fractionary density to $\Omega_{b0}\simeq 0.05$. 

The continuity equation for baryons reads

\begin{equation}
\Delta_b^{\prime} + a\theta_b - 3\psi^{\prime} = 0,
\end{equation}
and the Euler equation 

\begin{equation}
\left( a\theta_b \right) ^{\prime} +\mathcal{H}a \theta_b - k^2 \phi =0.
\end{equation}

Now, following a similar derivation as used previously to treat the dark matter fluid  and
applying again the quasi-static approximation, the Einstein equations become

\begin{equation}\label{Einstein0bqs}
-k^2 \psi  = \frac{3}{2} \mathcal{H}_0^2 a^2 
\left( \frac{\Omega_{b0}}{a^3} \Delta_b +  \Omega_v \Delta_v \right) ,
\end{equation}

\begin{equation}\label{Einstein1bqs}
-\frac{k^2}{2}(\phi - \psi) = \frac{3\mathcal{H}^2}{\rho}\eta \, \theta_v,
\end{equation}

while the continuity equation for baryons is
\begin{equation}
a\theta_b = - \Delta_b^{\prime} .
\end{equation}

{}From the background dynamics, we have that
\begin{equation}
\frac{a}{H} \frac{dH}{da} = - \frac{3}{2} \frac{H_0^2}{H^2}  
\left[ (1+\omega_v) \Omega_v  + \frac{\Omega_{b0}}{a^3} \right] ,
\end{equation}
and

\begin{equation}
a \frac{d\omega_v}{da} = 3\omega_v (1+\omega_v) \left[ 1- \nu - \frac{\Omega_v}{2}\frac{H_0^2}{H^2} \right] - \frac{3}{2} \frac{H_0^2}{H^2}\omega_v \frac{\Omega_{b0}}{a^3}.
\label{newform}
\end{equation}
Now, using Eq.~(\ref{newform}) in the Euler equation, the equation for the density perturbation 
contrast for the baryons, $\Delta_b\equiv \delta\rho_b/(\rho_v+\rho_b)$, becomes

\begin{align}
& a^2 \frac{d^2 \Delta_b}{da^2} + \Big{\lbrace} 3 - \frac{3}{2} \frac{H_0^2}{H^2} \left[ \Omega_v (1+\omega_v) + \frac{\Omega_{b0}}{a^3} \right] \Big{\rbrace} a \frac{d\Delta_b}{da} - \frac{3}{2} \frac{H_0^2}{H^2}\frac{\Omega_{b0}}{a^3}\Delta_b \nonumber \\ 
& = \Big{[} \frac{3}{2}\frac{H_0^2}{H^2} \Omega_v + \frac{2 \tilde{\eta} a}{3 H_0 \Omega_v (1+2\omega _v)} 
\Big{(} \frac{3 H \omega_v}{a} + \frac{H^2}{H_0^2} \frac{\tilde{\xi } \nu}{\Omega_v} \left(\frac{\Omega_v}{\Omega_{v0}}\right)^{\nu} \Big{)} \Big{]} \Delta_v 
\nonumber \\ 
& - \frac{2 \tilde{\eta} H a}{3 H_0 \Omega_v (1+2\omega)} \frac{d\Delta_v}{da}.
\label{Deltab}
\end{align}

The previous viscous fluid density perturbation equation~(\ref{euler}) 
is also modified when including baryons and it now becomes  

\begin{align}\label{eulerv}
&a^2 \frac{d^2 \Delta_v}{d a^2}+\left[3 - \frac{3}{2}\Omega_v \frac{H_0^2}{H^2} - \frac{3}{2}\frac{\Omega_{b0}}{a^3} \frac{H_0^2}{H^2} +\bar{A} +
    k^2 B \right] a \frac{d \Delta_v}{da} \nonumber \\ & + \left( \bar{C} + k^2 D \right) \Delta_v = \frac{3}{2} \frac{H_0^2}{H^2} \frac{\Omega_{b0}}{a^3} \frac{(1+2\omega_v)}{(1+\omega_v)}\Delta_b,
\end{align}
with $\Delta_v\equiv \delta\rho_v/(\rho_v+\rho_b)$ and where the factors $B$ and $D$ have the same form as defined before, Eqs.~(\ref{termB}) and (\ref{termD}), respectively,
while the factors $\bar{A}$ and $\bar{C}$ appearing in the above equation (\ref{eulerv}) 
are defined, respectively, as

\begin{eqnarray}
\bar{A} &=& A +   \frac{3 \omega_v}{2(1+2\omega_v)(1+\omega_v)} 
\frac{\Omega_{b0}}{a^3}\frac{H_0^2}{H^2},
\label{newtermA}
\end{eqnarray}
and

\begin{eqnarray}
\bar{C} &=& C + \frac{9 \omega_v (2+6\omega_v + 5\omega_v^2)}{2(1+2\omega_v)(1+\omega_v)} 
\frac{\Omega_{b0}}{a^3}\frac{H_0^2}{H^2}. 
\end{eqnarray}
We have now a two-fluid system described by the coupled equations 
(\ref{Deltab}) and (\ref{eulerv}) and where the baryon density contrast 
enters as a source term in the dark matter viscous equation one.

In the previous sections, we have analyzed the viscous dark matter overdensity growth 
(when in the absence of baryons) assuming $\Omega_{v0}=0.3$. When including the 
baryonic component, we now have the splitting $\Omega_{\rm matter}
\equiv \Omega_v +\Omega_b$, with $\Omega_{v0}=0.25$ and $\Omega_{b0}=0.05$.
Hence, one possible interpretation is that we have previously considered that even baryons were subjected to the viscous effects.

\begin{figure}[!htb]
\centering
\includegraphics[width=7.5cm]{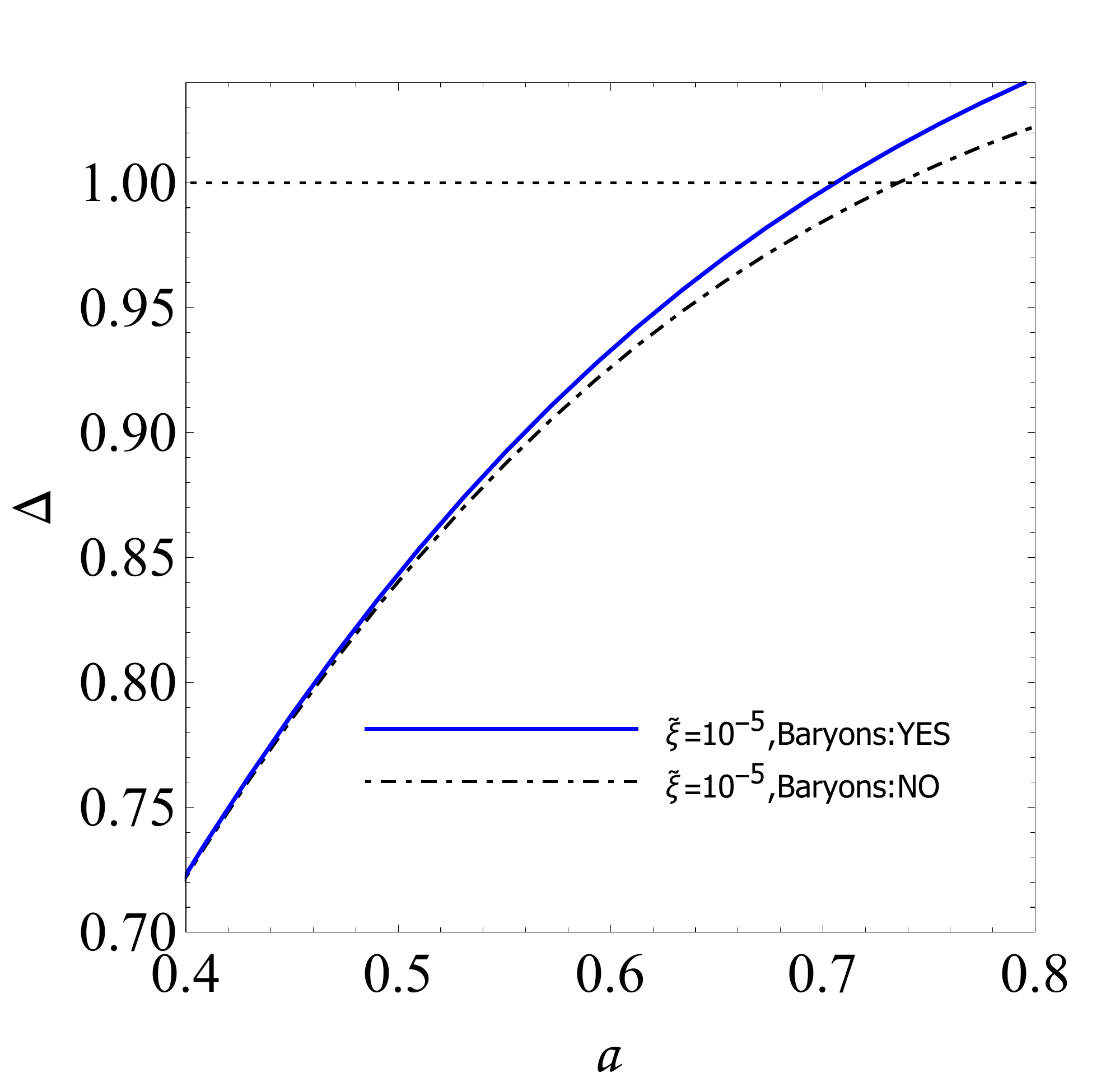}
\caption{The density contrast $\Delta$ as a function of the scale factor $a$ with $\tilde{\xi}=10^{-5}$, $\nu=\lambda=0$ and $R=1$. The solid line represents $\Delta_{eff}$ (where baryons are present). The dashed-dotted line is the same as in Panel (b) of figure \ref{k02ab}.}
\label{figbaryons}
\end{figure}

In figure \ref{figbaryons} we fix $\tilde{\xi}=10^{-5}$ and $R=1$ in the constant
viscosities case as an illustrative example of the effect of baryons on
the evolution of the effective 
total density contrast $\Delta_{\rm eff}$, defined as being the weight 
averaged quantity,

\begin{equation}
\Delta_{\rm eff}=\frac{\Omega_v\Delta_{v}+
\Omega_{b}\Delta_b}{\Omega_{v}+\Omega_{b}}.
\label{Deff}
\end{equation}
In figure \ref{figbaryons}, the dashed-dotted line corresponds to the case previously shown in 
panel (b) of figure~\ref{k02ab}, while the solid line corresponds to the case where baryons
are accounted for, following Eq.~(\ref{Deff}).
In both cases we notice that the influence of the background expansion 
is almost the same. This is simple because 
values of the viscosities of order $10^{-5}$ do not lead to a relevant 
deviation from the standard pressureless dark matter background scaling $\propto a^{-3}$,
at low values of the scale factor $a$. 
Nevertheless, we note that the inclusion of standard pressureless baryons make
the growth suppression in $\Delta_{\rm eff}$ to be not so efficient for a fixed viscosity
as it did in the previous case of absence of baryons. Thus, a higher value of the viscosity parameters are required such as to be able to lead to the same growth suppression as 
observed before. Therefore, we can conclude that the inclusion of baryons tends to lead 
to slightly different upper bounds on the dark matter viscosity.


\section{Conclusions}
\label{conclusions}

In this work we have studied the combined effects of both bulk and shear viscosities
on how they affect the perturbations relevant for structure formation. 
Our study is motivated by the observational evidence in favor of modifications
of the concordance $\Lambda$CDM model, which has difficulty in explaining a number of
important problems, such as
the excessive agglomeration of matter due
the nature of the CDM, the puzzle of missing satellites,
the cusp-core problem and also the issue related to
the fact that the Planck collaboration observed less clusters than
expected. In particular, the apparent excess of
clustered structures indicates the importance of processes able to suppress the 
density contrast growth, such as viscous effects. Previous studies on this issue
have included only effects due to the bulk viscosity and by assuming that the
dark matter fluid is endowed only with this viscous effect. Shear viscosity has always
been tacitly assumed to lead to negligible effects. This, at first, seems to
be a good assumption given that shear effects can only act at the level of
the perturbations for a homogeneous and isotropic universe described by the
FLRW metric, while bulk effects act already at the background level.
However, our results show that the shear can be as efficient as the bulk
in damping the perturbations and delaying the transition to the nonlinear regime.
In fact, we have seen that is physically more acceptable and also natural 
to have a shear viscosity acting concomitantly with the bulk viscosity, as far as their 
dependence on the cosmological fluid density is concerned. 

Our results obtained with the bulk viscosity and in the absence of the shear viscosity, 
they are found to be in agreement with previous ones~\cite{Velten:2013pra,Velten:2013rra,Velten:2014xca}, 
indicating that a bulk viscous dark matter can have the effect of alleviating
the excess of power existing in the standard cold dark matter scenario.
On the other hand, when the effect of the shear viscosity is considered, 
our results show a strengthening 
of the suppression of power at small scales (except for some special cases discussed in 
section~\ref{results}). 
This suppression is similar in many aspects to that verified in the warm dark matter 
scenario~\cite{warm}.  However, perturbations in the viscous fluid may exhibit 
important differences with respect to the warm dark matter case for some range of scales, mainly at very small scales. 
The construction of a complete realistic viscous model for the dark sector faces, however, important challenges as, 
for example, the analysis of deep nonlinear regime addressing the cusp-core problem in galaxies.
This requires numerical simulations taking into account
the viscous properties discussed here. Our results, thus, points to the importance of considering the possible
effects of the viscosities of the fluids considered in these simulations, such as to 
modeling a more physical and realistic situation that can be in effect in the structure
formation problem. Moreover, as a future follow-up of the present work, a more fundamental approach to the hydrodynamics formulation will be interesting to be analyzed, determining more precise forms for the viscous coefficients, and asking probably 
for a causal description for the viscous fluid, in the spirit of the M\"uller-Israel-Stewart formalism~\cite{muller,israel,stewart}.

As a consequence of our results on the effects of the viscous effects on the density contrast
and done in conjunction
with a statistical analysis, we have been also able to determine some upper bounds on the
overall magnitudes for the bulk and shear viscosities as 
$\xi \lesssim 1.427\times10^{-6} H_0/(24 \pi G) \sim 4.0 \times 10^{-12} {\rm GeV}^3 \simeq 58.6$Pa sec and 
$\eta \lesssim 2.593\time10^{-6} H_0/(24 \pi G)\sim 7.8 \times 10^{-12} {\rm GeV}^3 \simeq 106.5$Pa sec, 
respectively. It must be recalled, however, that these results are based on the much simpler situation where the
viscosities are constant. We have shown that varying viscosities with the fluid density can lead to a much
richer and varied possibilities. A detailed statistical analysis in this case is, on the other hand,
a difficulty task, but we hope to address this with more details elsewhere.
We also have shown that the inclusion of baryons, done in
the most conservative analysis where the baryons are taken as a pressureless fluid,
leads to looser upper bounds on the dark matter viscosity. 

{}Finally, since the viscous effects studied here can be expected to be associated with intrinsic 
properties of the
dark matter component of the universe, these bounds can eventually help to provide 
future constraints on dark matter candidates and in their searches, through the properties they might have,
e.g., their interactions. The study done in this work can be seen as a necessary initial 
study on the combined effects of both bulk and shear viscosities on structure formation and whose results 
can very well have other ramifications and to be of importance in unveiling the properties of dark matter.


\acknowledgments

We thank CNPq (Brazil) for partial financial support. J.C.F., H.V. and
C.S.M.B. thank also CAPES (Brazil) and FAPES (Brazil) for partial financial support.
R.\,O.\,R.~is partially supported by  Conselho Nacional de
Desenvolvimento Cient\'{\i}fico e Tecnol\'ogico - CNPq (Grant
No.~303377/2013-5) and Funda\c{c}\~ao Carlos Chagas Filho de Amparo
\`a Pesquisa do Estado do Rio de Janeiro - FAPERJ (Grant
No.~E-26/201.424/2014) and Coordena\c c\~ao de Pessoal de N\'{\i}vel Superior
- CAPES (Processo No. 88881.119017/2016-01).

\vspace{1cm}


\end{document}